\documentclass[a4paper,journal,11pt,draftclsnofoot,onecolumn]{IEEEtran}
\usepackage{amsmath}
\usepackage{amssymb}
\usepackage{graphicx}
\usepackage{cite}
\usepackage{color}
\usepackage{theorem}
\usepackage{url}
\usepackage{algorithmic}
\newtheorem{prop}{Proposition}
\theorembodyfont{\rmfamily}
\newtheorem{rem}{Remark}

\newcommand{\vect}[1]{\boldsymbol{#1}}
\newcommand{\mat}[1]{\boldsymbol{#1}}

\newcommand{\E}[1]{{\mathsf{E}}\left\{#1\right\}}

\newcommand{\Tr}[1]{{\mathrm{Tr}}\left\{#1\right\}}
\newcommand{\ex}[1]{\exp \left\{#1\right\}}
\newcommand{\range}[1]{\mathcal{R}\left(#1\right)}
\newcommand{\etr}[1]{{\mathrm{etr}}\left\{#1\right\}}
\newcommand{\diag}[1]{{\mathrm{diag}}\left(#1\right)}

\newcommand{\bc}{\vect{c}}
\newcommand{\bd}{\vect{d}}
\newcommand{\q}{\vect{q}}
\newcommand{\bs}{\vect{s}}
\newcommand{\x}{\vect{x}}
\newcommand{\y}{\vect{y}}
\newcommand{\bm}{\vect{m}}

\newcommand{\A}{\mat{A}}
\newcommand{\B}{\mat{B}}
\newcommand{\C}{\mat{C}}
\newcommand{\D}{\mat{D}}
\newcommand{\Dbar}{\bar{\mat{D}}}
\newcommand{\bE}{\mat{E}}
\newcommand{\F}{\mat{F}}
\newcommand{\Y}{\mat{Y}}

\newcommand{\I}{\mat{I}}
\newcommand{\M}{\mat{M}}
\newcommand{\N}{\mat{N}}
\newcommand{\Null}{\mat{Q}_{\perp}}
\newcommand{\bP}{\mat{P}}
\newcommand{\Pbar}{\bar{\mat{P}}}
\newcommand{\Q}{\mat{Q}}
\newcommand{\R}{\mat{R}}
\newcommand{\iR}{\mat{R}^{-1}}

\newcommand{\bS}{\mat{S}}

\newcommand{\bu}{\vect{u}}
\newcommand{\U}{\mat{U}}
\newcommand{\Ummsd}{\hat{\mat{U}}_{\text{mmsd}}}
\newcommand{\Ummsdlmbgh}{\hat{\mat{U}}_{\text{mmsd-LM-B}}}
\newcommand{\Ummsdlmvmf}{\hat{\mat{U}}_{\text{mmsd-LM-vMF}}}

\newcommand{\Uorth}{\mat{U}_{\perp}}

\newcommand{\Ubar}{\bar{\mat{U}}}
\newcommand{\Vbar}{\bar{\mat{V}}}

\newcommand{\X}{\mat{X}}
\newcommand{\z}{\vect{z}}
\newcommand{\Z}{\mat{Z}}

\newcommand{\bgamma}{\vect{\gamma}}
\newcommand{\bGamma}{\mat{\Gamma}}
\newcommand{\blambda}{\vect{\lambda}}
\newcommand{\bLambda}{\mat{\Lambda}}
\newcommand{\btheta}{\vect{\theta}}
\newcommand{\bthetammsd}{\hat{\vect{\theta}}_{\text{mmsd}}}

\newcommand{\varn}{\sigma_{n}^{2}}
\newcommand{\ivarn}{\sigma_{n}^{-2}}

\newcommand{\vMF}[1]{\mathrm{vMF}\left( #1\right)}  
\newcommand{\Bgh}[1]{\mathrm{B}\left( #1\right)} 
\newcommand{\BMF}[3]{\mathrm{BMF}\left( #1,#2,#3\right)} 
\newcommand{\mB}[4]{\tilde{\mathrm{B}}\left( #1,#2,#3,#4\right)} 
\newcommand{\vBMF}[2]{\mathrm{vBMF}\left( #1,#2\right)} 

\newcommand{\pdftG}[4]{\mathcal{G}_t\left(#1,#2,#3,#4\right)}

\newcommand{\oneFone}[3]{\thinspace_1 F_{1}\left(#1,#2;#3\right)}
\newcommand{\zeroFone}[2]{\thinspace_0 F_{1}\left(#1;#2\right)}

\begin{document}
\title{Minimum mean square distance \\estimation of a subspace}

\author{Olivier Besson$^{(1)}$, Nicolas Dobigeon$^{(2)}$ and
Jean-Yves Tourneret$^{(2)}$\\
\normalsize $^{(1)}$ University of Toulouse, ISAE, Department Electronics Optronics Signal, Toulouse, France\\
\normalsize $^{(2)}$ University of Toulouse, IRIT/INP-ENSEEIHT/T\'eSA, Toulouse, France\\
\small\texttt{olivier.besson@isae.fr,\{Nicolas.Dobigeon,Jean-Yves.Tourneret\}@enseeiht.fr}}
\maketitle
\begin{abstract}
We consider the problem of subspace estimation in a Bayesian setting. Since we are operating in the Grassmann manifold, the usual approach which consists of minimizing the mean square error (MSE) between the true subspace $\U$ and its estimate $\hat{\U}$ may not be adequate as the MSE is not the natural metric in the Grassmann manifold. As an alternative, we propose to carry out subspace estimation by minimizing the mean square distance (MSD) between $\U$ and its estimate, where the considered distance is a natural metric in the Grassmann manifold, viz. the distance between the projection matrices. We show that the resulting estimator is no longer the posterior mean of $\U$ but entails computing the principal eigenvectors of the posterior mean of $\U \U^{T}$.  Derivation of the MMSD estimator is carried out in a few illustrative examples including a linear Gaussian model for the data and a Bingham or von Mises Fisher prior distribution for $\U$. In all scenarios, posterior distributions are derived and the MMSD estimator is obtained either analytically or implemented via a Markov chain Monte Carlo simulation method. The method is shown to provide accurate estimates even when the number of samples is lower than the dimension of $\U$. An application to hyperspectral imagery is finally investigated.
\end{abstract}

\newpage
\section{Problem statement}
In many signal processing applications, the signals of interest do not span the entire observation space and a relevant and frequently used assumption is that they evolve in a low-dimensional subspace \cite{Scharf91}. Subspace modeling is accurate when the signals consist of a linear combination of $p$ modes in a $N$-dimensional space, and constitute a good approximation for example when the signal covariance matrix is close to rank-deficient. As a consequence, subspace estimation plays a central role in recovering these signals with maximum accuracy. An ubiquitous solution to this problem is to resort to the singular value decomposition (SVD) of the data matrix. The SVD emerges naturally as the maximum likelihood estimator in the classical model $\Y = \U \bS + \N$, where $\Y$ stands for the $N \times K$ observation matrix, $\U$ is the (deterministic) $N \times p$ matrix, with $p<N$, whose columns span the $p$-dimensional subspace of interest, $\bS$ is the $p \times K$ (deterministic) waveform matrix and $\N$ is the additive noise. The $p$ principal left singular vectors of $\Y$ provide very accurate estimates of a basis for the range space $\range{\U}$ of $\U$, and have been used successfully, e.g., in estimating the frequencies of damped exponentials or the directions of arrival of multiple plane waves, see \cite{Kumaresan82,Kumaresan83} among others. However, the SVD can incur some performance loss in two main cases, namely when the signal-to-noise ratio (SNR) is very low and thereof the probability of a subspace swap or subspace leakage is high \cite{Thomas95,Hawkes01,Johnson08,Nadakuditi10}. A second case occurs when the number of samples $K$ is lower than the subspace dimension $p$: indeed, $\Y$ is at most of rank $K$ and information is lacking about how to complement $\range{\Y}$ in order to estimate $\range{\U}$.

Under such circumstances, a Bayesian approach might be helpful as it
enables one to assist estimation by providing some statistical
information about $\U$. We investigate such an approach herein and
assign to the unknown matrix $\U$ an appropriate prior distribution,
taking into account the specific structure of $\U$. The paper is
organized as follows. In section \ref{section:MMSD}, we propose an
approach based on minimizing a natural distance on the Grassmann
manifold, which yields a new estimator of $\U$. The theory is
illustrated in section \ref{section:examples} where the new
estimator is derived for some specific examples. In section
\ref{section:simulations} its performance is assessed through
numerical simulations, and compared with conventional approaches.
Section \ref{section:appli} studies an application to the analysis
of interactions between pure materials contained in hyperspectral
images.

\section{Minimum mean square distance estimation \label{section:MMSD}}
In this section, we introduce an alternative to the conventional
minimum mean square error (MMSE) estimator, in the case where a
subspace is to be estimated. Let us consider that we wish to
estimate the range space of $\U$ from the joint distribution
$p\left(\Y,\U\right)$ where $\Y$ stands for the available data
matrix. Usually, one is not interested in $\U$ \emph{per se} but
rather in its range space $\range{\U}$,  and thus we are operating
in the Grassmann manifold $G_{N,p}$, i.e., the set of
$p$-dimensional subspaces in $\mathbb{R}^{N}$ \cite{Edelman98}. It
is thus natural to wonder whether the MMSE estimator (which is  the
chief systematic approach in Bayesian estimation \cite{Kay93}) is
suitable in $G_{N,p}$. The MMSE estimator $\hat{\btheta}$ of a
vector $\btheta$ minimizes the average Euclidean distance between
$\hat{\btheta}$ and $\btheta$, i.e., $\E{\left\| \hat{\btheta} -
\btheta \right\|_{2}^{2}}$. Despite the fact that this distance is
natural in an Euclidean space, it may not be the more natural metric
in $G_{N,p}$.  In fact, the natural distance between two subspaces
$\range{\U_1}$ and $\range{\U_2}$ is given by $\left(\sum_{k=1}^{p}
\theta_k^2\right)^{1/2}$ \cite{Edelman98} where $\theta_k$ are the
principal angles between these subspaces, which can be obtained by
SVD of $\U_2^T \U_1$ where $\U_1$ and $\U_2$ denote orthonormal
bases for these subspaces \cite{Golub96}. The SVD of $\U_2^T \U_1$
is defined as $\U_2^T \U_1 = \X \diag{\cos \theta_1 , \cdots, \cos
\theta_p} \Z^T$, where $\X$ and $\Z$ are two $p \times p$ unitary matrices. Therefore, it seems more adequate, rather than
minimizing $\left\| \hat{\U} - \U \right\|_{F}^{2}$ as the MMSE
estimator does, to minimize the natural distance between the
subspaces spanned by $\hat{\U}$ and $\U$. Although this is the most
intuitively appealing method, it faces the drawback that the cosines
of the angles and not the angles themselves emerge naturally from
the SVD. Therefore, we consider minimizing the sum of the squared
sine of the angles between $\hat{\U}$ and $\U$, since for small
$\theta_k$, $\sin \theta_k \simeq \theta_k$. As argued in
\cite{Edelman98,Golub96}, this cost function is natural in the
Grassmann manifold since it corresponds to the Frobenius norm of the
difference between the projection matrices on the two subspaces, viz
$\sum_{k=1}^{p} \sin^2 \theta_k = \left\| \hat{\U} \hat{\U}^{T} - \U
\U^{T} \right\|_{F}^{2} \triangleq d^{2}\left( \hat{\U},\U \right)$.
It should be mentioned that our approach follows along the same
principles as in \cite{Srivastava00} where a Bayesian framework is
proposed for subspace estimation, and where the author considers
minimizing  $d\left( \hat{\U},\U \right)$. Hence the theory
presented in this section is similar to that of
\cite{Srivastava00}, with some exceptions. Indeed the
parameterization of the problem in \cite{Srivastava00} differs from
ours and the application of the theory is also very different, see
the next section.

Given that $d^{2}\left( \hat{\U},\U \right)=
2\left( p- \Tr{\hat{\U}^T \U \U^T \hat{\U}} \right)$, we define the
minimum mean-square distance (MMSD) estimator of $\U$  as
\begin{equation}
\Ummsd = \arg \max_{\hat{\U}} \E{\Tr{\hat{\U}^T \U \U^T \hat{\U}}}.
\end{equation}
Since
\begin{equation}
\E{\Tr{\hat{\U}^T \U \U^T \hat{\U}}} = \\ \int \left[ \int \Tr{\hat{\U}^T \U \U^T \hat{\U}} p\left(\U | \Y \right) d \U \right] p\left(\Y\right) d \Y
\end{equation}
it follows that
\begin{align}
\Ummsd &= \arg \max_{\hat{\U}} \int \Tr{\hat{\U}^T \U \U^T \hat{\U}} p\left(\U | \Y \right) d \U \nonumber \\
&= \arg \max_{\hat{\U}} \Tr{\hat{\U}^T \left[ \int \U \U^T p\left(\U | \Y \right) d \U \right] \hat{\U} }.
\end{align}
Therefore, the MMSD estimate of the subspace spanned by $\U$ is
given by the $p$ largest eigenvectors of the matrix $\int \U \U^T
p\left(\U | \Y \right) d \U$, which we denote as
\begin{equation}\label{Ummsd}
\Ummsd = \mathcal{P}_{p}\left\{ \int \U \U^T p\left(\U | \Y \right) d \U \right\}.
\end{equation}
In other words, MMSD estimation amounts to find the best rank-$p$
approximation to the posterior mean of the projection matrix $\U
\U^T$ on $\range{\U}$. For notational convenience, let us denote
$\M\left(\Y\right)=\int \U \U^T p\left(\U | \Y \right) d \U$. Except
for a few cases where this matrix can be derived in closed-form (an
example is given in the next section), there usually does not exist
any analytical expression for $\M\left(\Y\right)$. In such situation, an efficient way
to approximate the matrix $\M\left(\Y\right)$ is to use a Markov chain Monte Carlo simulation
method whose goal is to generate random matrices $\U$ drawn from the
posterior distribution $ p\left(\U | \Y \right)$, and to approximate
the integral in \eqref{Ummsd} by a finite sum. This aspect will be
further elaborated in the next section. Let $\M\left(\Y\right)=
\U_{M}(\Y) \mat{L}_{M}(\Y) \U_{M}^{T}(\Y)$ denote the eigenvalue
decomposition of $\M\left(\Y\right)$ with $\mat{L}_{M}(\Y) =
\diag{\ell_{1}(\Y),\ell_{2}(\Y),\cdots,\ell_{N}(\Y)}$ and
$\ell_{1}(\Y) \geq \ell_{2}(\Y) \geq\cdots \geq \ell_{N}(\Y)$.  Then
the average distance between $\Ummsd$ and $\U$ is given by
\begin{align}\label{HSbound}
\E{d^{2} \left( \Ummsd, \U \right)} &= 2p-2\int \left[ \int \Tr{\Ummsd^T \U \U^T \Ummsd} p\left(\U | \Y \right) d \U \right] p\left(\Y\right) d \Y \nonumber \\
&=2p-2\int \Tr{\Ummsd^{T} \M\left(\Y\right) \Ummsd} p\left(\Y\right) d \Y \nonumber \\
&= 2p  - 2 \sum_{k=1}^{p} \int \ell_{k}(\Y) p\left(\Y\right) d \Y.
\end{align}
The latter expression constitutes a lower bound on $\E{d^{2}\left(
\hat{\U}, \U \right)}$ and is referred to as the Hilbert-Schmidt
bound in \cite{Srivastava00,Grenander98}. As indicated in these
references, and similarly to $\M\left(\Y\right)$, this lower bound
may be difficult to obtain analytically.

The MMSD approach can be extended to the mixed case where, in
addition to $\U$, a parameter vector $\btheta$ which can take
arbitrary values in $\mathbb{R}^{q}$ needs to be estimated jointly
with $\U$. Under such circumstances, one can estimate $\U$ and
$\btheta$ as
\begin{align}
\left( \Ummsd , \bthetammsd \right) = \arg \min_{\hat{\U},
\hat{\btheta}} \mathsf{E} \left\{ -\Tr{\hat{\U}^T \U \U^T \hat{\U}}
+\left(\hat{\btheta}-\btheta\right)^{T}
\left(\hat{\btheta}-\btheta\right) \right\}.
\end{align}
Doing so, the MMSD estimator of $\U$ is still be given by
\eqref{Ummsd} while the MMSD and MMSE estimators of $\btheta$
coincide.
\begin{rem}\label{rem:MMSD_vs_MSME}
The MMSD approach differs from an MMSE approach which would entail
calculating the posterior mean of $\U$, viz $\int \U p\left(\U | \Y
\right) d \U$.  Note that the latter may not be meaningful, in
particular when the posterior distribution $p\left(\U | \Y \right)$
depends on $\U$ only through $\U \U^T$, see next section for an
example. In such a case, post-multiplication of $\U$ by any $p
\times p$ unitary matrix $\Q$ yields the same value of  $p\left(\U |
\Y \right)$. Therefore averaging $\U$ over $p\left(\U | \Y \right)$
does not make sense while computing \eqref{Ummsd} is relevant. On
the other hand, if $p\left(\U | \Y \right)$ depends on $\U$
directly, then computing the posterior mean of $\U$ can be
investigated: an example where this situation occurs will be
presented in the next section. As  a final comment,  observe that
$\int \U p\left(\U | \Y \right) d \U$ is not necessarily unitary but
its range space can be used to estimate $\range{\U}$.
\end{rem}
\begin{rem}
We open a parenthesis here regarding the framework of this paper.
Although it is not directly related to this paper (we do not address
optimization problems here) it is interesting to note the recent
growing interest in optimization problems on special manifolds,
especially on the Stiefel manifold (the set of $N \times p$
matrices $\U$ such that $\U^T \U = \I$) and the Grassmann manifold,
see the excellent tutorial paper by Edelman \emph{et al.}
\cite{Edelman98} as well as \cite{Absil08,Absil09}, and
\cite{Abrudan08,Abrudan09,Fiori09b} for signal processing
applications. These references show the interest of taking into
account the underlying geometry of the problem, as we attempt to do
herein.
\end{rem}

\section{Illustration examples \label{section:examples}}
In this section we illustrate the previous theory on some examples,
including the conventional  linear Gaussian model (conditioned on
$\U$) and a model involving the eigenvalue decomposition of the data
covariance matrix. As a first step, we address the issue of
selecting prior distributions for $\U$ and then move on to the
derivation of the MMSD estimator.
\subsection{Prior distributions}
\begin{figure}[htb]
\centering
\includegraphics[width=8cm]{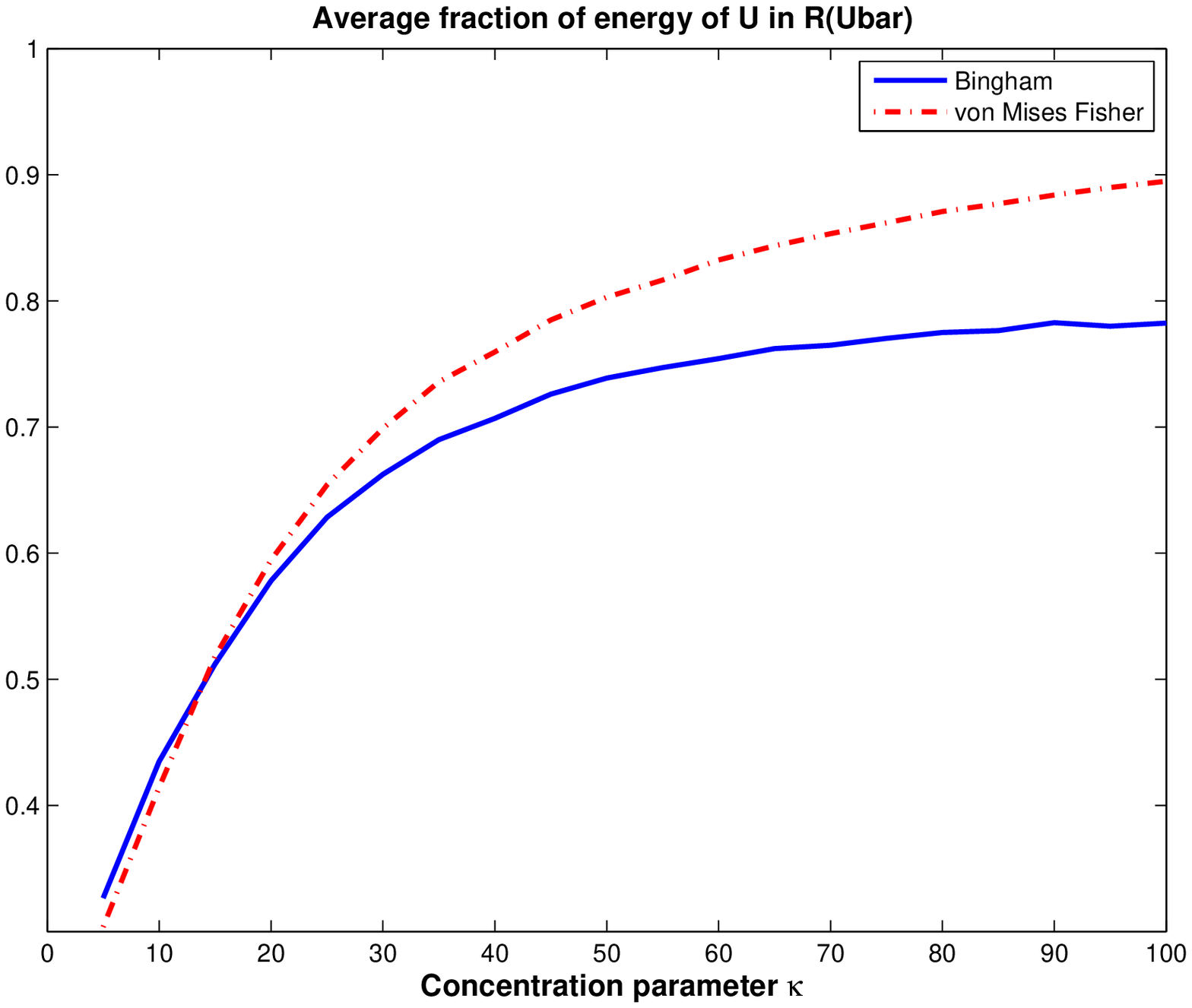}
\caption{Average fraction of energy of $\U$ in $\range{\Ubar}$ versus $\kappa$. $N=20$, $p=5$.}
\label{fig:mean_energy}
\end{figure}

\begin{figure}[htb]
\centering
\includegraphics[width=8cm]{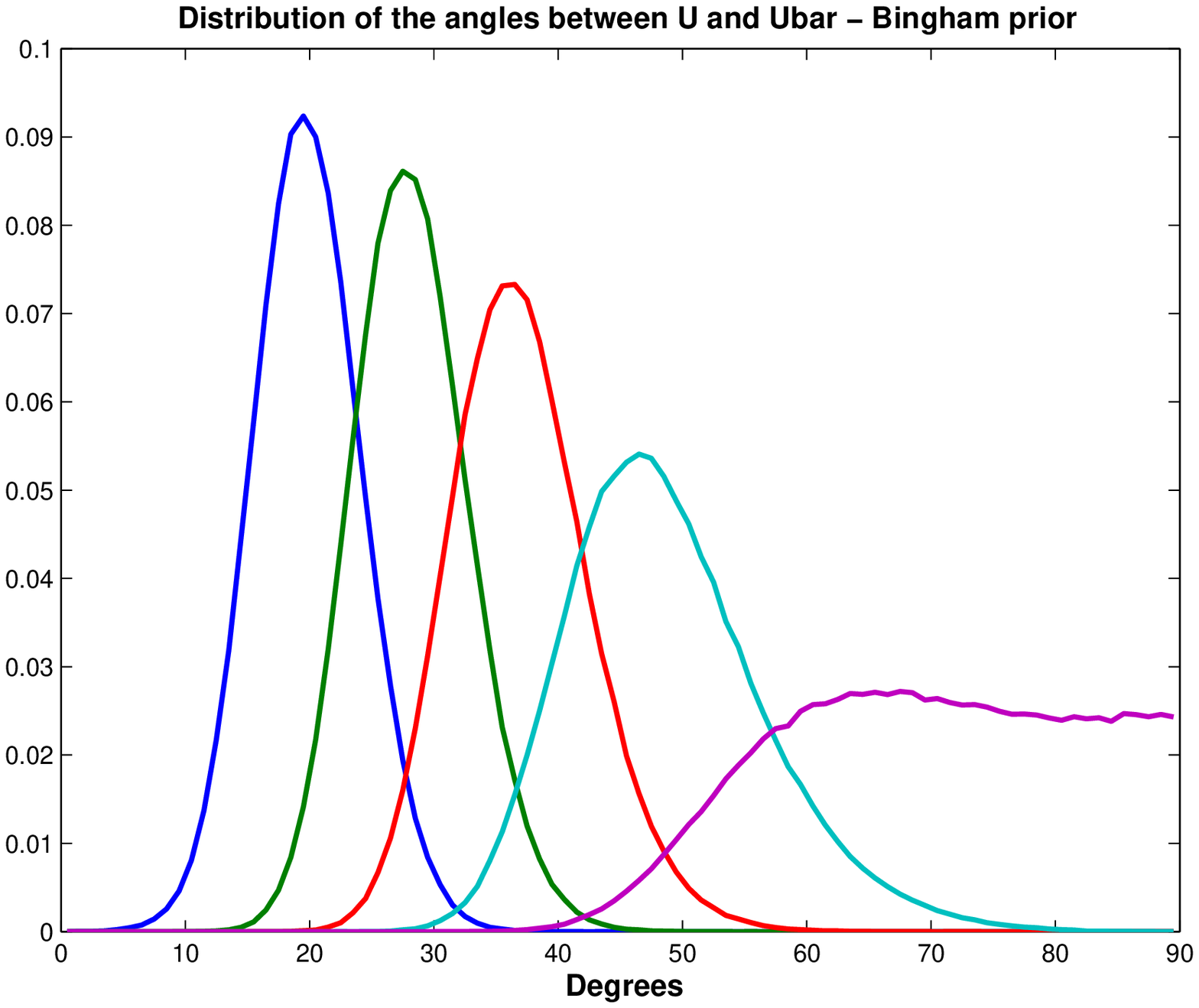}
\caption{Distribution of the angles between $\range{\U}$ and $\range{\Ubar}$ for a Bingham distribution. $N=20$, $p=5$ and $\kappa=20$.}
\label{fig:distrib_angle_bingham}
\end{figure}

\begin{figure}[htb]
\centering
\includegraphics[width=8cm]{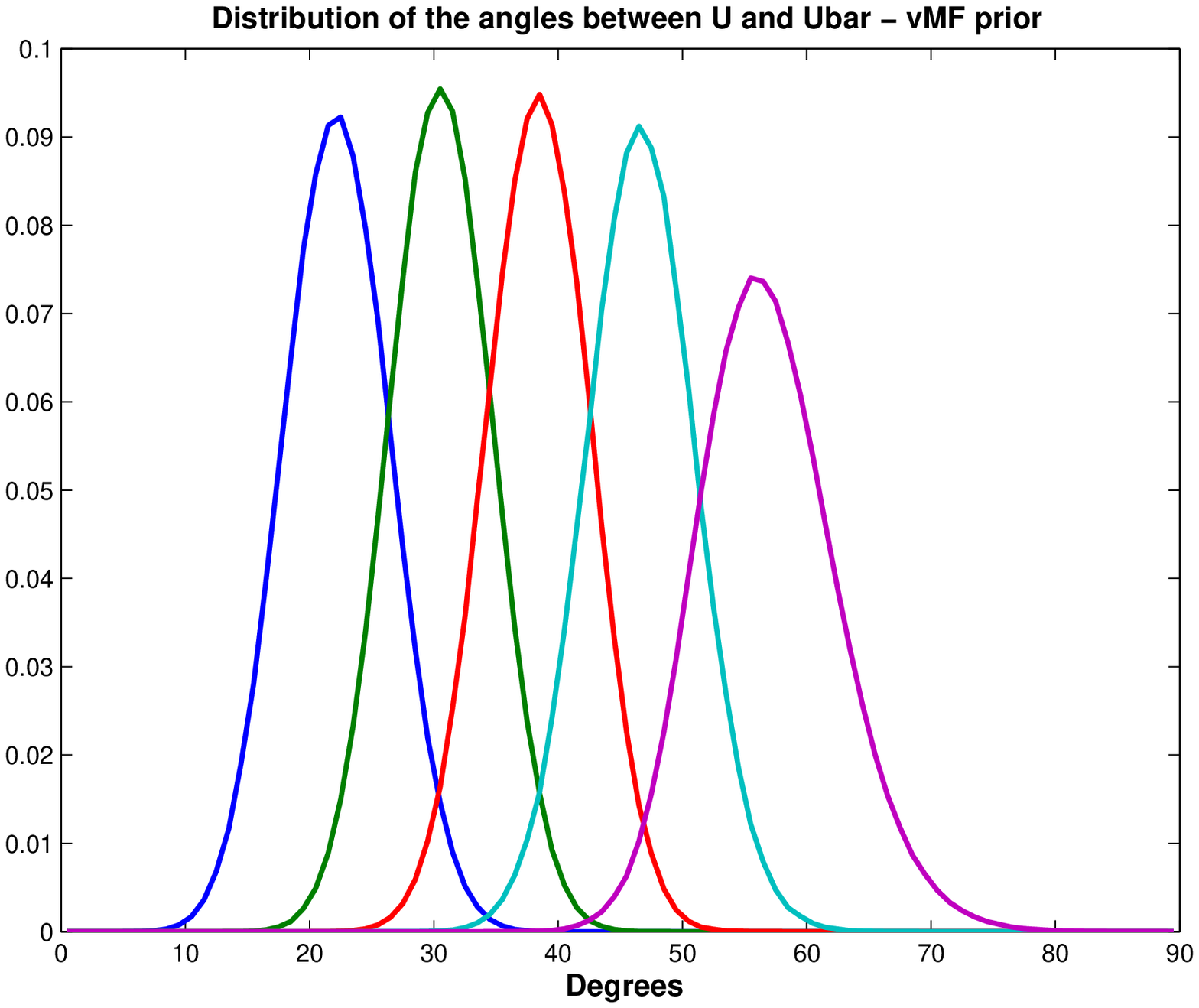}
\caption{Distribution of the angles between $\range{\U}$ and $\range{\Ubar}$ for a von Mises Fisher distribution. $N=20$, $p=5$ and $\kappa=20$.}
\label{fig:distrib_angle_vmf}
\end{figure}

A crucial step in any Bayesian estimation scheme consists of
selecting the prior distribution for the variables to be estimated.
We focus here on distributions on the Stiefel or Grassmann manifold,
depending whether we consider the matrix $\U$ itself or its range
space. There exist only a few distributions on the Stiefel or
Grassmann manifolds, the most widely accepted being the Bingham or
von Mises Fisher (vMF) distributions \cite{Mardia99,Chikuse03},
which are given respectively by
\begin{align}
p_{\mathrm{B}} (\U)  &=\frac{1}{\oneFone{\frac{1}{2}p}{\frac{1}{2}N}{\A}}  \etr{\U^{T} \A \U} \\
p_{\mathrm{vMF}} (\U) & = \frac{1}{\zeroFone{\frac{1}{2}N}{\frac{1}{4}\F^{T}\F}} \etr{\F^{T} \U}
\end{align}
where   $\etr{.}$ stands for the exponential of the trace of the
matrix between braces,  $\A$ is an $N \times N$ symmetric matrix,
$\F$ is an $N \times p$ arbitrary matrix, and $\zeroFone{a}{\X}$,
$\oneFone{a}{b}{\X}$ are hypergeometric functions of matrix
arguments, see e.g. \cite{Chikuse03} for their definitions. We will
denote these distributions as $\Bgh{\A}$ and $\vMF{\F}$,
respectively. Observe that the Bingham distribution depends on $\U
\U^{T}$ only, and can thus be viewed as a distribution on the
Grassmann manifold \cite{Mardia99,Chikuse03} while the vMF
distribution depends on $\U$ and is a distribution on the Stiefel
manifold. In our case, in order to introduce some knowledge about
$\U$, we assume that it is ``close'' to a given subspace spanned by
the columns of an orthonormal matrix $\Ubar$, and hence we consider
two possible prior distributions for $\U$, namely
\begin{align}
\pi_{\mathrm{B}} \left(\U\right) &\propto \etr{\kappa \U^T \Ubar \Ubar^T \U} \label{Bingham} \\
\pi_{\mathrm{vMF}} \left(\U\right) &\propto \etr{\kappa \U^T \Ubar} \label{vMF}
\end{align}
where  $ \propto$ means ``proportional to''. The distribution in
\eqref{Bingham} is proportional to the sum of the squared cosine
angles between $\range{\U}$ and $\range{\Ubar}$ while
$\pi_{\mathrm{vMF}} \left(\U\right)$ is proportional to the sum of
the cosine angles between $\range{\U}$ and $\range{\Ubar}$. Note that $\kappa$
is a concentration parameter: the larger $\kappa$ the more
concentrated around $\Ubar$ are the subspaces $\U$. The difference
between the two distributions is the following. In the Bingham
distribution only $\range{\U}$ and $\range{\Ubar}$ are close (at
least for large values of $\kappa$) since $\pi_{\mathrm{B}}\left(\U\right)$ is
invariant to post-multiplication of $\U$ by any $p \times p$ unitary
matrix $\Q$ . Hence $\U$ is not necessarily close to $\Ubar$. In
contrast, under the vMF prior distribution, $\U$ and $\Ubar$ are
close. For illustration purposes, Figure \ref{fig:mean_energy}
displays the average fraction of energy of $\U$ in $\range{\Ubar}$ defined as
\begin{equation}
\mathrm{AFE}\left( \U, \Ubar \right) = \E{\Tr{\U^T \Ubar \Ubar^T \U}/p}.
\end{equation}
As can be observed from these figures, both distributions allow the distance between $\U$ and $\Ubar$ to be
set in a rather flexible way.
Their AFE is shown to be identical for small values of the
concentration parameter but, when $\kappa$ increases, the AFE of the
vMF distribution increases faster.
Additionally, even if the AFE are close for small values of $\kappa$, the
distributions of the angles between $\range{\U}$ and $\range{\Ubar}$
exhibit some differences, as shown in Figures
\ref{fig:distrib_angle_bingham} and \ref{fig:distrib_angle_vmf}
which display the probability density functions of these angles for
$\kappa=20$.

\subsection{Linear model}
In order to illustrate how the previous theory can be used in
practice, we first consider a simple example, namely a linear
Gaussian model (conditioned on $\U$), i.e., we assume that the data
follows the model $\Y = \U \bS + \N$ where the columns of $\N$ are
independent and identically distributed (i.i.d.) Gaussian vectors
with zero-mean and (known) covariance matrix $\varn \I$. We assume
that no knowledge about $\bS$ is available and hence its prior
distribution is set to $\pi \left( \bS \right) \propto 1$.
Therefore, conditioned on $\U$ we have
\begin{align}\label{p(Y|U)_LM}
&p\left(\Y | \U\right) = \int p\left(\Y | \U, \bS \right) \pi \left( \bS \right) d \bS \nonumber \\
&\propto \int \etr{-\frac{1}{2\varn} \left(\Y-\U\bS\right)^{T} \left(\Y-\U\bS\right)} d \bS \nonumber \\
&\propto \etr{-\frac{1}{2\varn} \Y^T \Y + \frac{1}{2\varn} \Y^T \U \U^T \Y}.
\end{align}
When $\U$ follows the Bingham prior distribution, the posterior
distribution of $\U$, conditioned on $\Y$ is given by
\begin{equation}\label{p(U|Y)_LM_Bingham}
p\left(\U | \Y\right) \propto \etr{ \U^T \left[ \kappa \Ubar \Ubar^T + \frac{1}{2\varn} \Y \Y^T \right] \U}
\end{equation}
which is recognized as a Bingham distribution with parameter matrix
$\kappa \Ubar \Ubar^T + \frac{1}{2\varn} \Y \Y^T $, i.e., $\U | \Y
\sim \Bgh{\kappa \Ubar \Ubar^T + \frac{1}{2\varn} \Y \Y^T }$. For such a Bingham distribution, it turns out that the
eigenvectors of $\int \U \U^T p\left(\U | \Y \right) d \U$ coincide with those of $\kappa \Ubar \Ubar^T +
\frac{1}{2\varn} \Y \Y^T$, with the same ordering of their eigenvalues, see Appendix \ref{app:eigBingham} for a proof. Therefore the MMSD estimator is obtained in \textbf{closed-form} as
\begin{equation}\label{Ummsd_LM_Bingham}
\Ummsdlmbgh = \mathcal{P}_{p} \left\{ \kappa \Ubar \Ubar^T + \frac{1}{2\varn} \Y \Y^T \right\}.
\end{equation}
Therefore, the MMSD estimator has a very simple form in this case.
It consists of the principal subspace of a (weighted) combination of
the a priori projection matrix $\Ubar \Ubar^T$ and the information
brought by the data through $\Y \Y^T$. Observe that, in this
particular case of a Bingham posterior, the MMSD estimator coincides
with the maximum a posteriori (MAP) estimator.

Let us now consider the case where the prior distribution of $\U$ is
vMF, and contrast it with the previous example. Using
\eqref{p(Y|U)_LM} along with along with \eqref{vMF}, it follows that
the posterior distribution now writes
\begin{equation}\label{p(U|Y)_LM_vMF}
p\left(\U | \Y\right) \propto \etr{  \kappa \U^T \Ubar + \frac{1}{2\varn} \U^T \Y \Y^T \U}
\end{equation}
which is referred to as the Bingham-von-Mises-Fisher (BMF)
distribution  with parameter matrices $\Y \Y^T$, $\frac{1}{2\varn}
\I$ and $\kappa \Ubar$ respectively\footnote{The matrix $\X$ is said
to have a $\BMF{\A}{\B}{\C}$ distribution -where $\A$ is an $N\times
N$ symmetric matrix, $\B$ is a $p\times p$ diagonal matrix and $\C$
is an $N \times p$ matrix- if  $p(\X) \propto \etr{\C^T \X + \B \X^T
\A \X}$.}. Although this distribution is known \cite{Chikuse03}, to our knowledge,
there does not exist any analytic expression for the integral in
\eqref{Ummsd} when $\U | \Y$  has the BMF distribution \eqref{p(U|Y)_LM_vMF}. Therefore, the MMSD estimator cannot be
computed in closed-form. In order to remedy this problem, a Markov
chain Monte Carlo simulation method can be advocated
\cite{Robert04,Robert07} to generate a large number of matrices
$\U^{(n)}$ drawn from \eqref{p(U|Y)_LM_vMF}, and to approximate
\eqref{Ummsd} as
\begin{equation}\label{Uiam}
\Ummsdlmvmf \simeq \mathcal{P}_{p}\left\{ \frac{1}{N_r} \sum_{n=N_{\textrm{bi}}+1}^{N_{\textrm{bi}}+N_r} \U^{(n)} \U^{(n)^H} \right\}.
\end{equation}
In \eqref{Uiam}, $N_{\textrm{bi}}$ is the number of burn-in samples
and $N_{r}$ is the number of samples used to approximate the
estimator. An efficient Gibbs sampling scheme to generate random
unitary matrices drawn from a $\BMF{\A}{\B}{\C}$ distribution with
arbitrary full-rank matrix $\A$ was proposed in \cite{Hoff09}. It
amounts to sampling successively each column of $\U$ by generating a
random unit norm vector drawn from a (vector) BMF distribution. In
our case, $\A=\Y \Y^T$ whose rank is $\min(K,N)$ and hence $\A$ is
rank-deficient whenever $K < N$, a case of most interest to us. Note
also that to generate matrices $\U$ drawn from the Bingham
distribution in \eqref{Bingham}, we need to consider $\A=\Ubar
\Ubar^T$ which has rank $p<N$. Therefore, the scheme of
\cite{Hoff09} needs to be adapted in order to generate random
matrices drawn from \eqref{p(U|Y)_LM_vMF}. In Appendix \ref{app:BMF}, we
review the method of \cite{Hoff09} and show how it can be modified
to handle the case of a rank-deficient matrix $\A$.
\begin{rem}\label{rem:iam}
Interestingly enough, the above estimator in \eqref{Uiam} is the
so-called induced arithmetic mean (IAM) \cite{Sarlette09} of the set
of unitary matrices $\U^{(n)}$, $n=N_{\textrm{bi}}+1,\cdots,N_{\textrm{bi}}+N_{r}$. It
differs from the Karcher mean of the set $\U^{(n)}$,
$n=N_{\textrm{bi}}+1,\cdots,N_{\textrm{bi}}+N_{r}$, which truly minimizes the sum of
the distances to all $\U^{(n)}$. However, the Karcher mean may not
exist and requires iterative schemes to be computed
\cite{Begelfor06} while the IAM is straightforward to compute.
\end{rem}
\begin{rem}\label{rem:map_lm_bingham}
In the particular case where $\U$ has a Bingham prior distribution,
the MAP estimator of $\U$ and its MMSD estimator are equal. This is
no longer true when $\U$ has a vMF prior distribution, and hence a
BMF posterior distribution. The mode of the latter is not known in
closed-form either. However, it can be approximated by selecting,
among the matrices generated by the Gibbs sampler, the matrix which
results in the largest value of the posterior distribution.
\end{rem}

\subsection{Covariance matrix model}
We now consider a more complicated case where $\Y$, conditioned on
$\U$ and $\bLambda$, is Gaussian distributed with zero-mean and
covariance matrix
\begin{equation}
\R = \E{\Y \Y^T} = \U \bLambda \U^{T} + \varn  \I
\end{equation}
where $\U$ is an orthonormal basis for the signal subspace,
$\bLambda$ is the diagonal matrix of the eigenvalues and $\varn$
stands for the white noise power which is assumed to be known here.
As it will be more convenient and more intuitively appealing, we
re-parametrize the covariance matrix as follows. The inverse of $\R$
can be written as
\begin{align}
\iR &= \U \left[ \left( \bLambda + \varn \I \right)^{-1} - \ivarn \I \right] \U^{T} + \ivarn \I \nonumber \\
&= \ivarn \I - \ivarn \U \bLambda \left( \bLambda + \varn \I \right)^{-1} \U^{T} \nonumber \\
&\triangleq \nu \I - \nu \U \left( \I- \bGamma \right) \U^{T}
\end{align}
where $\nu \triangleq \ivarn$, $\bGamma \triangleq \diag{\bgamma}$ with $\bgamma =\begin{bmatrix} \gamma_1 &\gamma_2 &\cdots &\gamma_p \end{bmatrix}^{T}$ and
\begin{equation}
0 < \gamma_k \triangleq \frac{\varn}{\varn+\lambda_k} < 1.
\end{equation}
The idea is to parametrize the problem in terms of $\U$ and
$\bGamma$ rather than $\U$ and $\bLambda$. The interest of this
transformation is twofold. First, it enables one to express all
eigenvalues with respect to the white noise level. Indeed, one has
$\R = \nu^{-1} \Uorth \Uorth^{T} + \nu^{-1} \U \bGamma^{-1} \U^{T}$
where $\Uorth$ is an orthonormal basis for $\range{\U}^{\perp}$ and
hence the $\gamma_k$s are representative of the scaling between the
``signal'' eigenvalues and the noise eigenvalues. In fact, they
carry information about the signal-to-noise ratio since
$\gamma_k=\left( 1 + \frac{\lambda_k}{\sigma^{2}} \right)^{-1}$ and
$\frac{\lambda_k}{\sigma^{2}}$ represents the SNR of the $k$-th
signal component. Second, this new parametrization will facilitate
derivation of the conditional distributions required for the Gibbs
sampler.

Since $\Y$ conditioned on $\U$ and $\bgamma$ is Gaussian, it follows that
\begin{equation}
p\left( \Y | \U, \bgamma  \right) = \left(2\pi\right)^{-NK/2} \left| \R \right|^{-K/2} \etr{-\frac{1}{2} \Y^{T} \iR \Y}.
\end{equation}
From $\iR = \nu \Uorth \Uorth^{T} + \nu \U \bGamma \U^{T}$, it
ensues that $\left| \iR \right| = \nu^{N} \left| \bGamma \right|$
and hence
\begin{equation}\label{p(Y|Ugamma)_ULU}
p\left( \Y | \U, \bgamma  \right) \propto  \left| \bGamma \right|^{K/2} \etr{-\frac{1}{2} \Y^{T} \left[ \nu \I - \nu \U \left( \I- \bGamma \right) \U^{T} \right] \Y}.
\end{equation}
Let us now consider the prior distributions for $\U$ and $\bgamma$ .
We will consider either a Bingham or vMF distribution for $\U$. As
for $\bgamma$, we assume that $\gamma_k$ are a priori independent
random variables uniformly distributed in the interval
$\left[\gamma_{-},\gamma_{+}\right]$, i.e.,
\begin{equation}\label{p(gamma)}
\pi (\bgamma) =\prod_{k=1}^{p} \left( \gamma_{+}-\gamma_{-} \right)^{-1} \mathbb{I}_{\left[\gamma_{-},\gamma_{+}\right]}(\gamma_k).
\end{equation}
The value of $\gamma_{+}$ [respectively $\gamma_{-}$] can be set to
$1$ [respectively $0$] if a non-informative prior is desired.
Otherwise, if some information is available about the SNR,
$\gamma_{-}$ and $\gamma_{+}$ can be chosen so as to reflect this
knowledge since $\gamma_k=\left( 1 + SNR_{k} \right)^{-1}$:
$\gamma_{+}$ [resp. $\gamma_{-}$] rules the lowest [resp. highest]
value of the SNR, say $SNR_{-}$ [resp. $SNR_{+}$].

With the Bingham assumption for $\pi\left(\U\right)$, the joint
posterior distribution of $\U$ and $\bgamma$ is
\begin{align}\label{joint_posterior_bingham}
&p\left( \U, \bgamma | \Y \right) \propto p\left( \Y | \U, \bgamma \right) \pi \left( \U \right) \pi \left(\bgamma\right)   \nonumber \\
&\propto \left| \bGamma \right|^{K/2} \left( \prod_{k=1}^{p} \mathbb{I}_{\left[\gamma_{-},\gamma_{+}\right]}(\gamma_k)\right) \nonumber \\
&\times \etr{\kappa \U^{T} \Ubar \Ubar^{T} \U +\frac{\nu}{2} \Y^{T} \U \left( \I- \bGamma \right) \U^{T} \Y}.
\end{align}
In order to come up with the posterior distribution of $\U$ only, we
need to marginalize \eqref{joint_posterior_bingham} with respect to
$\bgamma$. Let $\Z = \Y^{T} \U = \begin{bmatrix} \z_1 & \z_2 &
\cdots & \z_p \end{bmatrix}$. Then, from
\eqref{joint_posterior_bingham} one has
\begin{align}
p\left( \U | \Y \right)  &= \int p\left( \U, \bgamma | \Y \right)  d \, \bgamma \nonumber \\
& \propto   \etr{\kappa \U^{T} \Ubar \Ubar^{T} \U +\frac{\nu}{2} \U^{T} \Y \Y^{T} \U} \nonumber \\
&\times \prod_{k=1}^{p} \int_{\gamma_{-}}^{\gamma_{+}} \gamma_k^{K/2} \ex{- \frac{\nu}{2} \gamma_k \left\| \z_k \right\|^{2}} d \gamma_k \nonumber \\
& \propto \etr{\kappa \U^{T} \Ubar \Ubar^{T} \U +\frac{\nu}{2} \U^{T} \Y \Y^{T} \U} \nonumber \\
&\times \prod_{k=1}^{p} \left\| \z_k \right\|^{-2(1+K/2)} \left[ \gamma \left( \frac{\nu}{2} \gamma_{+} \left\| \z_k \right\|^{2}, 1+\frac{K}{2} \right) - \gamma \left( \frac{\nu}{2} \gamma_{-} \left\| \z_k \right\|^{2}, 1+\frac{K}{2} \right)\right]
\end{align}
where $\gamma(x,a)=\int_0^x t^{a-1} e^{-t} dt$ is the incomplete
Gamma function. Unfortunately, the above distribution does not
belong to any known family and it is thus problematic to generate
samples drawn from it. Instead, in order to sample according to
\eqref{joint_posterior_bingham}, we propose to use a Gibbs sampler
drawing samples according to $p\left( \U | \Y, \bgamma  \right)$ and
$p\left( \gamma_k | \Y, \U\right)$  for $k=1,\cdots,p$. From
\eqref{joint_posterior_bingham}, the conditional distribution of
$\U$ is
\begin{equation}\label{p(U|Ygamma)_bingham}
p\left( \U | \Y, \bgamma  \right) \propto \etr{\kappa \U^{T} \Ubar \Ubar^{T} \U + \frac{\nu}{2} \left( \I- \bGamma \right) \U^{T} \Y  \Y^{T} \U}
\end{equation}
which is recognized as a (modified) Bingham distribution\footnote{$\X \sim \mB{\A_1}{\B_1}{\A_2}{\B_2} \Leftrightarrow p(\X) \propto \etr{\B_1 \X^T \A_1 \X + \B_2 \X^T \A_2 \X}$}
\begin{equation}
\U | \Y, \bgamma \sim \mB{\Ubar \Ubar^{T}}{\kappa \I}{ \Y \Y^{T}}{\frac{\nu}{2} \left( \I- \bGamma \right)}.
\end{equation}
Let us now turn to the conditional distribution of $\bgamma | \Y,
\U$. From \eqref{joint_posterior_bingham} one has
\begin{align}\label{p(gamma|Uy)_bingham}
&p\left( \bgamma | \Y, \U \right) \propto \left| \bGamma \right|^{K/2} \etr{-\frac{\nu}{2} \Z \bGamma \Z^{T}}  \left( \prod_{k=1}^{p} \mathbb{I}_{\left[\gamma_{-},\gamma_{+}\right]}(\gamma_k)\right) \nonumber \\
&\propto \prod_{k=1}^{p} \left[ \gamma_{k}^{K/2} \ex{- \frac{\nu}{2} \left\| \z_{k} \right\|^{2} \gamma_{k}} \mathbb{I}_{\left[\gamma_{-},\gamma_{+}\right]}(\gamma_{k}) \right]
\end{align}
which is the product of independent gamma distributions with
parameters $\frac{K}{2}+1$ and $ \frac{\nu}{2} \left\| \z_k
\right\|^{2}$,  truncated in the interval
$\left[\gamma_{-},\gamma_{+}\right]$. We denote this distribution as
$\gamma_k \sim \pdftG{\frac{K}{2}+1}{\frac{\nu}{2} \left\| \z_k
\right\|^{2}}{\gamma_{-}}{\gamma_{+}}$. Random variables with such a
distribution can be efficiently generated using the accept-reject
scheme of \cite{Chung98}.

The above conditional distributions can now be used in a Gibbs
sampler, as described in Table  \ref{table:gibbs}. When $\U$ has a
vMF prior distribution, it is straightforward to show that $\U$,
conditioned on $\Y$ and $\bgamma$, follows a BMF distribution $\U |
\Y,\bgamma \sim \BMF{\Y \Y^{T}}{\frac{\nu}{2}\left( \I- \bGamma
\right) }{\kappa \Ubar}$ while the posterior distribution of
$\bgamma | \Y,\U$ is still given by \eqref{p(gamma|Uy)_bingham}.
Therefore line \ref{line:U|Ygamma} of the Gibbs sampler in Table
\ref{table:gibbs}  just needs to be modified in order to handle this
case.

\bigskip

\begin{table}[htb]
\begin{algorithmic}[1]
\REQUIRE initial values $\U^{(0)}$,  $\bgamma^{(0)}$
\FOR{$n=1,\cdots,N_{bi}+N_{r}$} \STATE sample $\U^{(n)}$ from
$\mB{\kappa \I}{\Ubar \Ubar^{T}}{\frac{\nu}{2} \left( \I-
\bGamma^{(n-1)} \right)}{ \Y \Y^{T}}$ in
\eqref{p(U|Ygamma)_bingham}. \label{line:U|Ygamma} \STATE for
$k=1,\cdots,p$, sample $\gamma_k^{(n)}$ from
$\pdftG{\frac{K}{2}+1}{\frac{\nu}{2} \left\| \Y^{T} \bu_k^{(n)}
\right\|^{2}}{\gamma_{-}}{\gamma_{+}}$ in
\eqref{p(gamma|Uy)_bingham}. \ENDFOR \ENSURE sequence of random
variables $\U^{(n)}$ and $\bgamma^{(n)}$
\end{algorithmic}
\caption{Gibbs sampler} \label{table:gibbs}
\end{table}

\section{Simulations \label{section:simulations}}
In this section we illustrate the performance of the approach
developed above through Monte Carlo simulations. In all simulations
$N=20$, $p=5$ and $\kappa=20$. The matrix $\bS$ is generated from a
Gaussian distribution with zero-mean and covariance matrix
$\sigma_{s}^{2} \I$ and the signal-to-noise ratio is defined as
$SNR=10  \log_{10} \left( \sigma_{s}^{2} / \varn \right)$. The
matrix $\U$ is generated from the Bingham distribution
\eqref{Bingham} or the vMF distribution \eqref{vMF} and, for the
sake of simplicity, $\Ubar = \begin{bmatrix} \I_p & \mat{0}
\end{bmatrix}^{T}$. The number of burn-in iterations in the Gibbs
sampler is set to $N_{\textrm{bi}}=10$ and $N_r=1000$. The MMSD estimator
\eqref{Ummsd} is compared with the MAP estimator, the MMSE
estimator, the usual SVD-based estimator and the estimator
$\hat{\U}=\Ubar$ that discards the available data and use only the a
priori knowledge. The latter is referred to as ``Ubar'' in the figures.
The estimators are evaluated in terms of the fraction of energy of
$\hat{\U}$ in $\range{\U}$, i.e., $\mathrm{AFE}\left( \hat{\U},\U
\right)$.
\subsection{Linear model}
\begin{figure}[p]
\centering
\includegraphics[width=8cm]{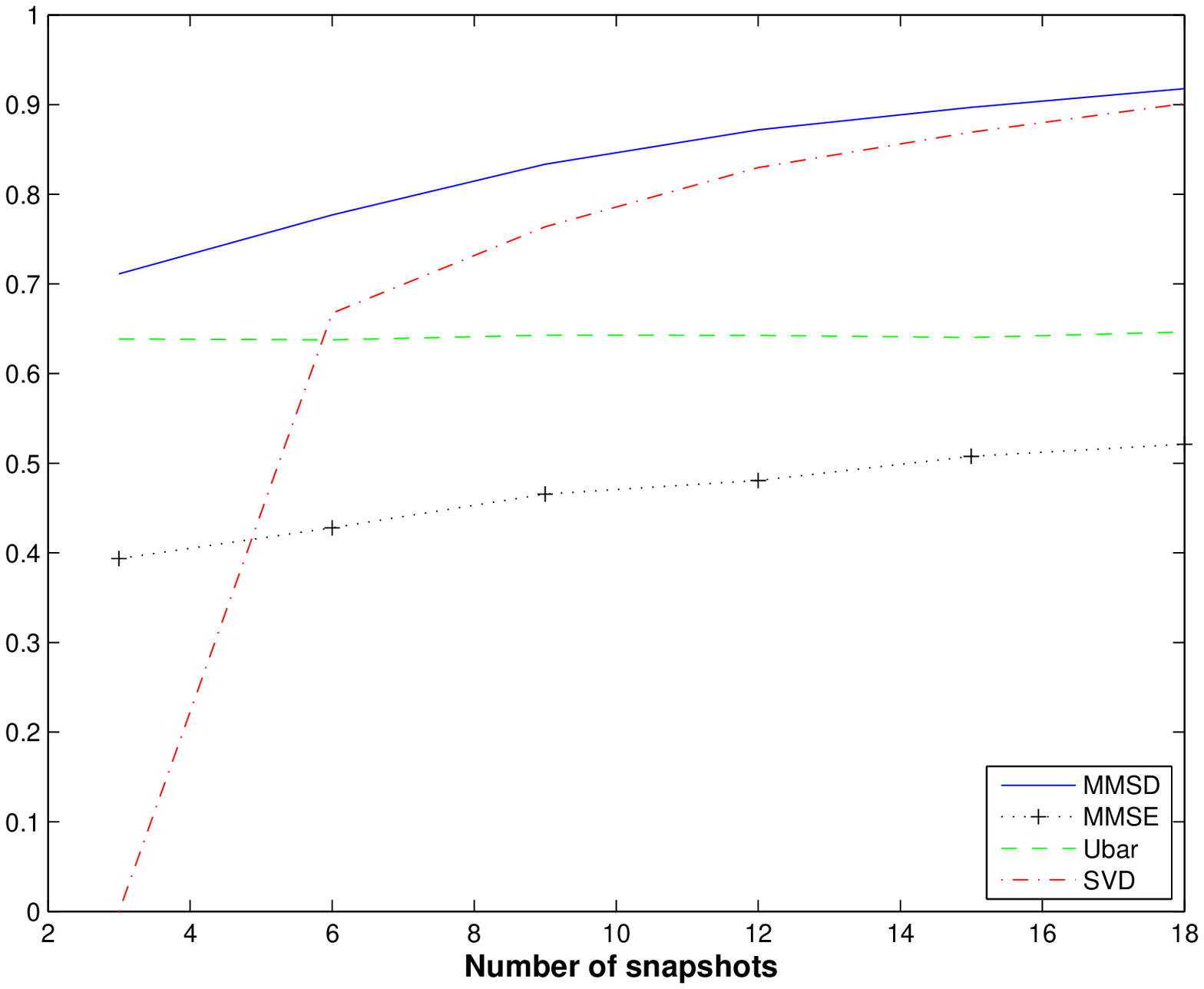}
\caption{Fraction of energy of $\hat{\U}$ in $\range{\U}$ versus $K$. $N=20$, $p=5$, $\kappa=20$ and $SNR=5$dB. Linear model,  Bingham prior.}
\label{fig:mmsd_lm_bingham_vs_K_N=20_p=5_kappa=20_SNR=5}
\end{figure}

\begin{figure}[p]
\centering
\includegraphics[width=8cm]{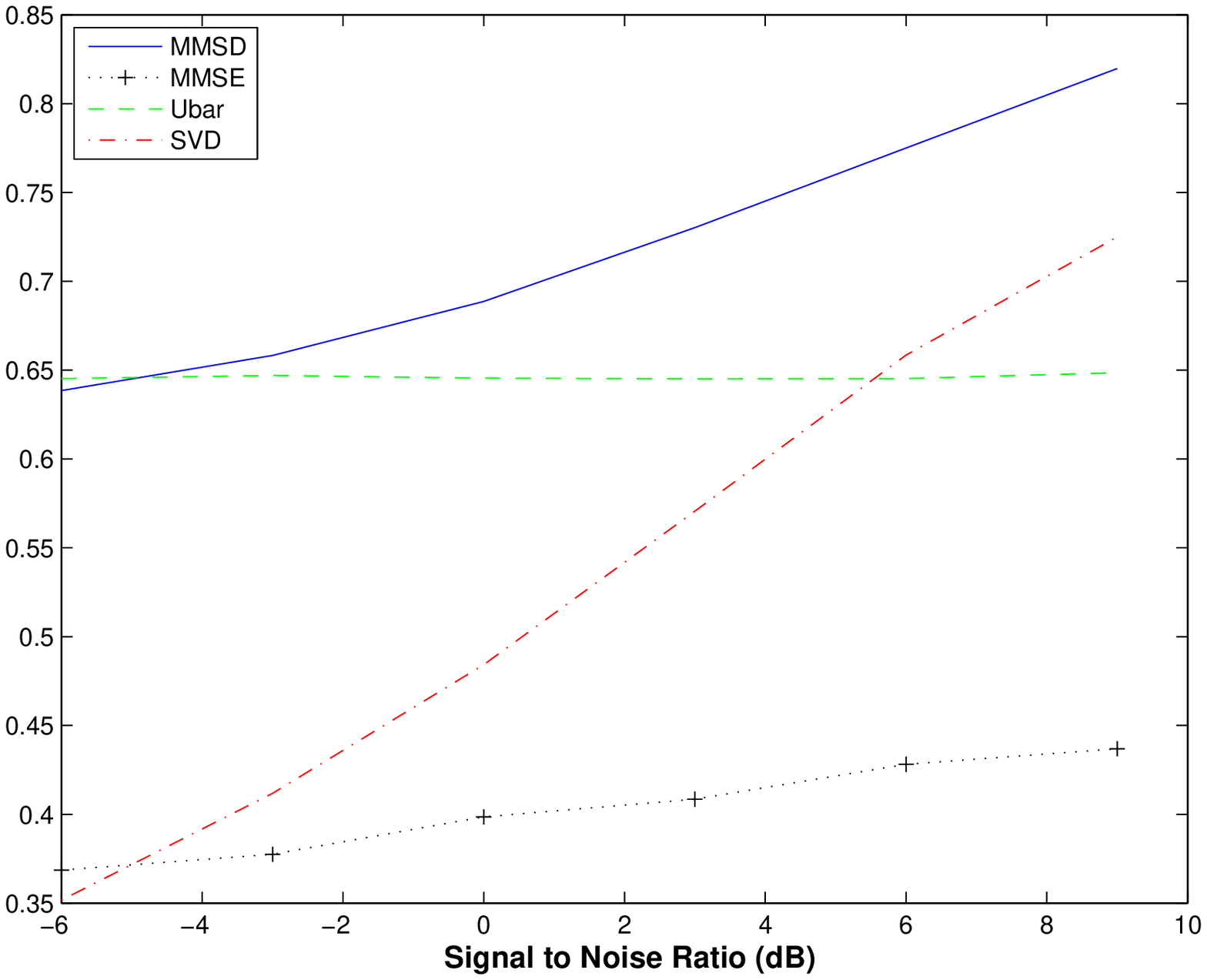}
\caption{Fraction of energy of $\hat{\U}$ in $\range{\U}$ versus $SNR$. $N=20$, $p=5$, $\kappa=20$ and $K=5$. Linear model,  Bingham prior.}
\label{fig:mmsd_lm_bingham_vs_SNR_N=20_p=5_kappa=20_K=5}
\end{figure}

\begin{figure}[p]
\centering
\includegraphics[width=8cm]{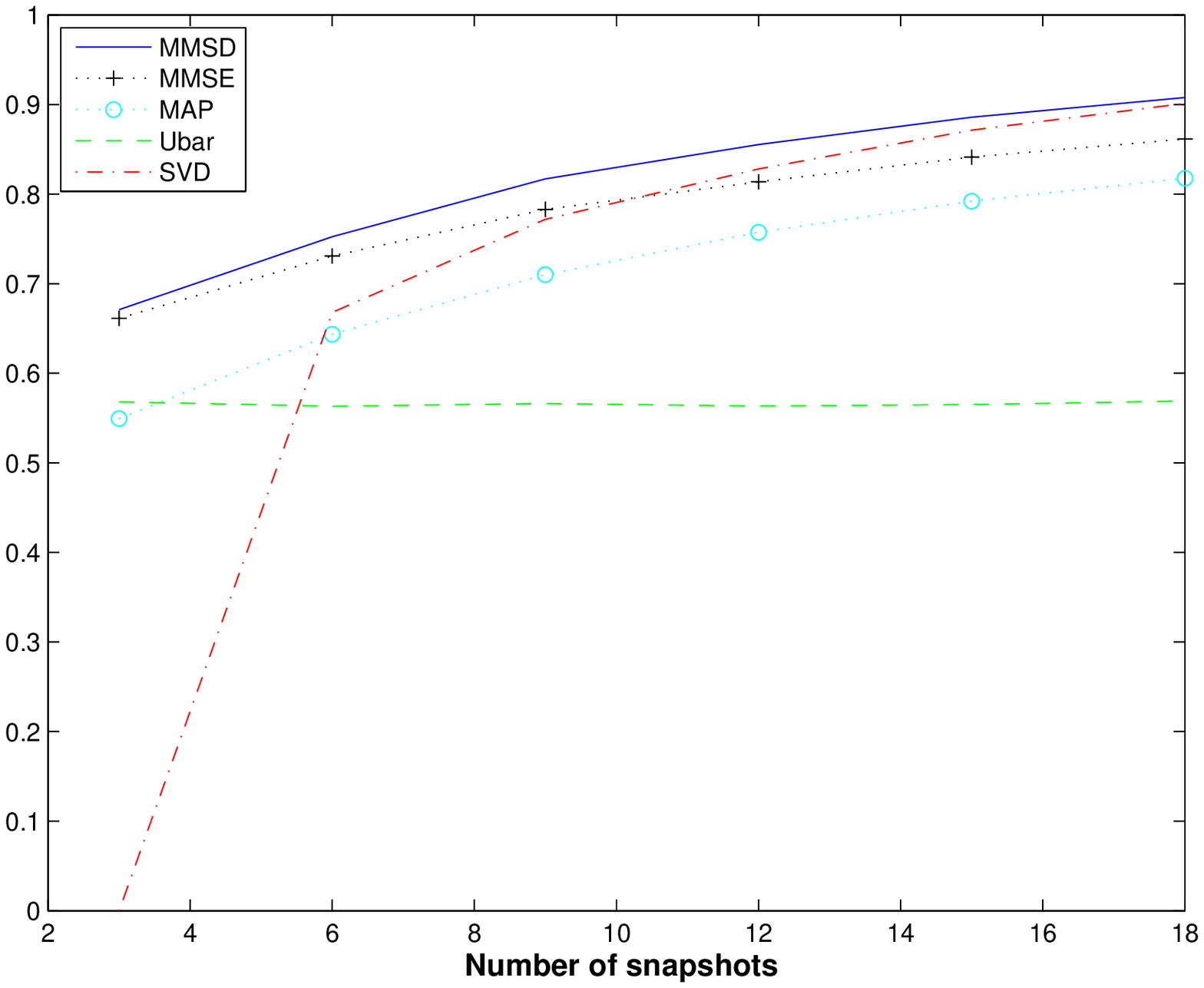}
\caption{Fraction of energy of $\hat{\U}$ in $\range{\U}$ versus $K$. $N=20$, $p=5$, $\kappa=20$ and $SNR=5$dB. Linear model,  vMF prior.}
\label{fig:mmsd_lm_vmf_vs_K_N=20_p=5_kappa=20_SNR=5}
\end{figure}

\begin{figure}[p]
\centering
\includegraphics[width=8cm]{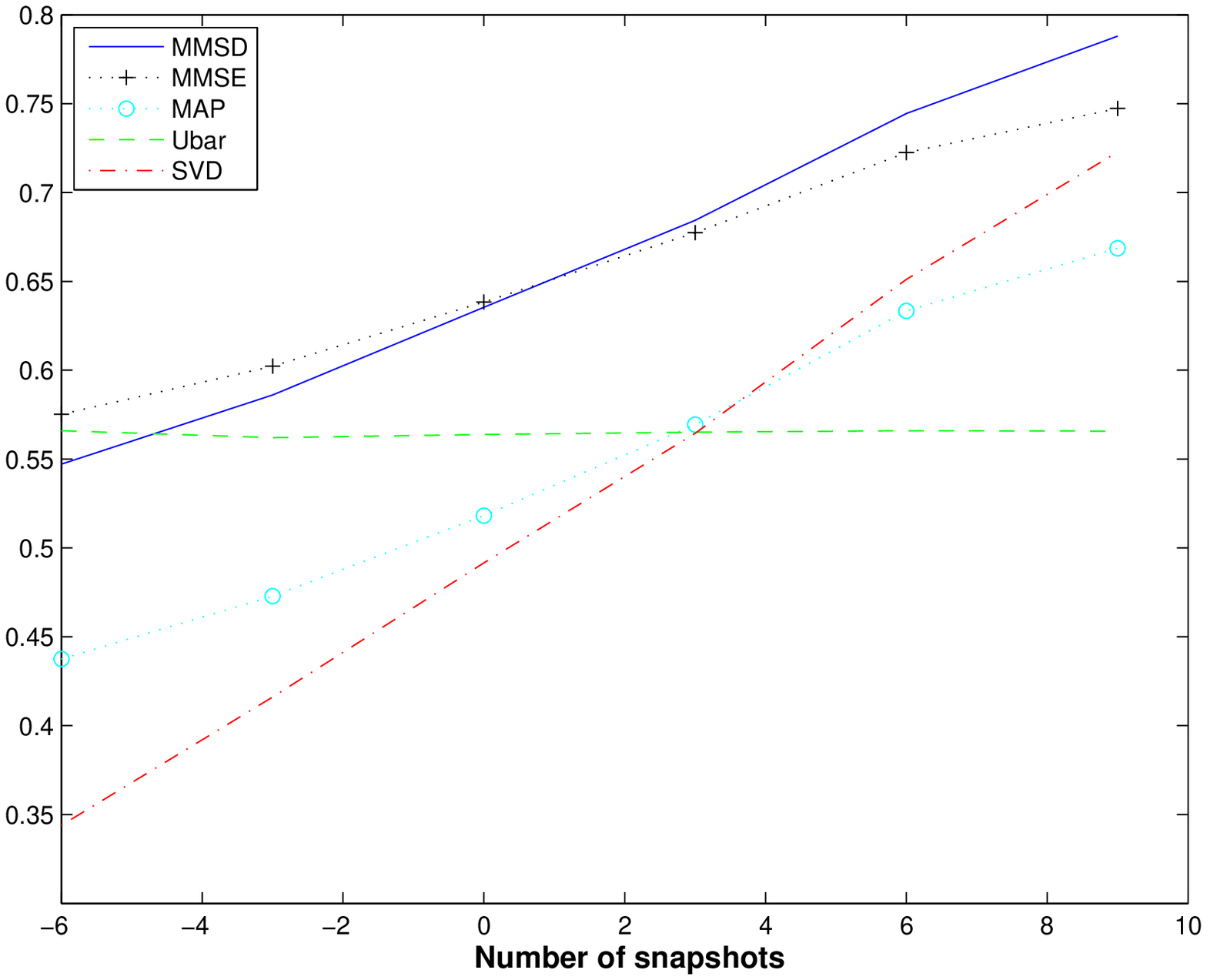}
\caption{Fraction of energy of $\hat{\U}$ in $\range{\U}$ versus $SNR$. $N=20$, $p=5$, $\kappa=20$ and $K=5$. Linear model,  vMF prior.}
\label{fig:mmsd_lm_vmf_vs_SNR_N=20_p=5_kappa=20_K=5}
\end{figure}

We begin with the linear model. Figures
\ref{fig:mmsd_lm_bingham_vs_K_N=20_p=5_kappa=20_SNR=5} to
\ref{fig:mmsd_lm_vmf_vs_SNR_N=20_p=5_kappa=20_K=5} investigate the
influence of $K$ and $SNR$ onto the performance of the estimators.
Figures \ref{fig:mmsd_lm_bingham_vs_K_N=20_p=5_kappa=20_SNR=5} and
\ref{fig:mmsd_lm_bingham_vs_SNR_N=20_p=5_kappa=20_K=5} concern the
Bingham prior while the vMF prior has been used to obtain Figures
\ref{fig:mmsd_lm_vmf_vs_K_N=20_p=5_kappa=20_SNR=5} and
\ref{fig:mmsd_lm_vmf_vs_SNR_N=20_p=5_kappa=20_K=5}. From inspection
of these figures, the following conclusions can be drawn:
\begin{itemize}
\item the MMSD estimator performs better than the estimator $\hat{\U}=\Ubar$, even at low SNR. The improvement is all the more pronounced that $K$ is large. Therefore, the MMSD estimator makes a sound use of the data to improve accuracy compared to using the prior knowledge only.
\item the MMSD estimator performs better than the SVD, especially at low SNR. Moreover, and this is a distinctive feature of this Bayesian approach, it enables one to estimate the subspace even when the number of snapshots $K$ is less than the size of the subspace $p$.
\item for a Bingham prior, the MMSE performs very poorly since the posterior distribution of $\U$ conditioned on $\Y$ depends on $\U \U^T$ only. Hence, averaging the matrix $\U$ itself does not make sense, see our remark \ref{rem:MMSD_vs_MSME}. In contrast, when $\U$ has a vMF prior, the posterior depends on both $\U$ and $\U \U^T$: in this case, the MMSE performs well and is close to the MMSD. Note however that the vMF prior is more restrictive than the Bingham prior.
\item the MMSD estimator also outperforms the MAP estimator.
\end{itemize}
As a conclusion, the MMSD estimator performs better than most other estimators in the large majority of cases.

\subsection{Covariance matrix model}
\begin{figure}[p]
\centering
\includegraphics[width=8cm]{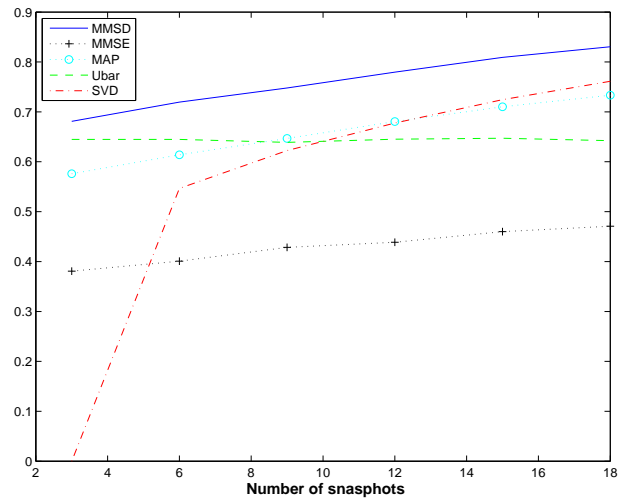}
\caption{Fraction of energy of $\hat{\U}$ in $\range{\U}$ versus
$K$. $N=20$, $p=5$, $\kappa=20$, $SNR_{-}=5$dB and $SNR_{+}=10$dB.
Covariance matrix model, Bingham prior.}
\label{fig:mmsd_ulu_bignham_vs_K_N=20_p=5_kappa=20_SNR=5-10}
\end{figure}

\begin{figure}[p]
\centering
\includegraphics[width=8cm]{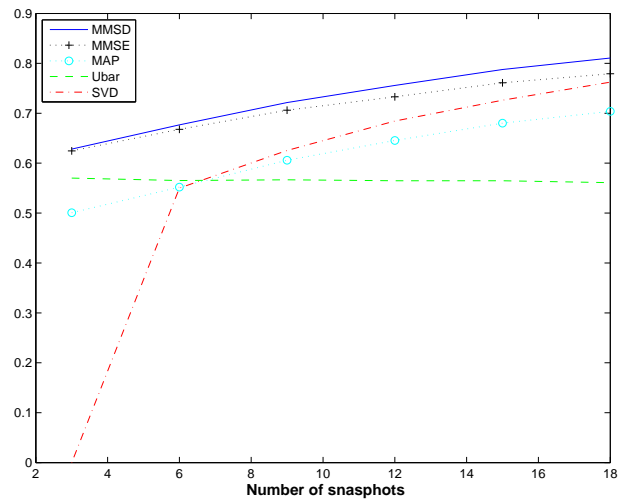}
\caption{Fraction of energy of $\hat{\U}$ in $\range{\U}$ versus
$K$. $N=20$, $p=5$, $\kappa=20$, $SNR_{-}=5$dB and $SNR_{+}=10$dB.
Covariance matrix model, vMF prior.}
\label{fig:mmsd_ulu_vmf_vs_K_N=20_p=5_kappa=20_SNR=5-10}
\end{figure}

We now conduct simulations with the covariance matrix model. The
simulation parameters are essentially the same as in the previous
section, except for the SNR. More precisely, the random variables
$\gamma_k$ are drawn from the uniform distribution in
\eqref{p(gamma)} where $\gamma_{-}$ and $\gamma_{+}$ are selected
such that $SNR_{-}=5$dB and $SNR_{+}=10$dB. The results are shown in
Fig. \ref{fig:mmsd_ulu_bignham_vs_K_N=20_p=5_kappa=20_SNR=5-10}
for the Bingham prior and Fig.
\ref{fig:mmsd_ulu_vmf_vs_K_N=20_p=5_kappa=20_SNR=5-10} for the vMF
prior. They corroborate the previous observations made on the linear
model, viz that the MMSD estimator offers the best performance over
all methods.

\section{Application to hyperspectral imagery} \label{section:appli}
In this section, we show how the proposed subspace estimation
procedure can be efficiently used for an application to multi-band
image analysis. For several decades, hyperspectral imagery has
received considerable attention because of its great interest for
various purposes: agriculture monitoring, mineral mapping, military
concerns, etc. One of the crucial issue when analyzing such image is
the spectral unmixing which aims to decompose an observed pixel
$\y_\ell$ into a collection of $R=p+1$ reference signatures,
$\bm_1,\ldots,\bm_R$ (called \emph{endmembers}) and to
retrieve the respective proportions of these signatures (or
\emph{abundances}) $a_{1,\ell},\ldots,a_{R,\ell}$ in this pixel \cite{Keshava2002}. To describe the
physical process that links the endmembers and their abundances to
the measurements, the most widely admitted mixing model is linear
\begin{equation}\label{eq:LMM}
\y_\ell = \sum_{r=1}^{R} a_{r,\ell} \bm_r
\end{equation}
where $\y_\ell \in \mathbb{R}^N$ is the pixel spectrum measured
in $N$ spectral bands, $\bm_r \in \mathbb{R}^N$
($r=1,\ldots,R$) are the $R$ endmember spectra and $a_{r,\ell}$
($r=1,\ldots,R$) are their corresponding abundances. Due to obvious
physical considerations, the abundances obey two kinds of
constraints. Since they represent proportions, they must satisfy the
following positivity and additivity constraints
\begin{equation} \label{eq:constraints}
\begin{cases}
a_{r,\ell} \geq 0,\quad r=1,\ldots,R,\\
\sum_{r=1}^R a_{r,\ell} =1.
\end{cases}
\end{equation}
Let now consider $L$ pixels $\y_1,\ldots,\y_L$ of an hyperspectral
image induced by the linear mixing model (LMM) in \eqref{eq:LMM}
with the abundance constraints \eqref{eq:constraints}. It is clear
that the dataset formed by these $L$ pixels lies in a
lower-dimensional subspace $\mathcal{U} \subset \mathbb{R}^{p}$.
More precisely, in this subspace $\mathcal{U}$, the dataset belongs
to a simplex whose vertices are the endmembers $\bm_1,\ldots,\bm_R$
to be recovered. Most of the unmixing strategies developed in the
hyperspectral imagery literature are based on this underlying
geometrical formulation of the LMM. Indeed, the estimation of the
endmembers is generally conducted in the lower-dimensional space
$\mathcal{U}$, previously identified by a standard dimension
reduction technique such as the principal component analysis (PCA)
\cite{Keshava2002}. However, it is well known that the model
linearity is a simplifying assumption and does not hold anymore in
several contexts, circumventing the standard unmixing algorithms.
Specifically, non-linearities are known to occur for scenes
including mixtures of minerals or vegetation. As a consequence,
evaluating the suitability of the LMM assumption for a given
hyperspectral image is a capital question that can be conveniently
addressed by the approach introduced above.

\subsection{Synthetic data}

\begin{figure}[p]
\centering
\includegraphics[width=8cm]{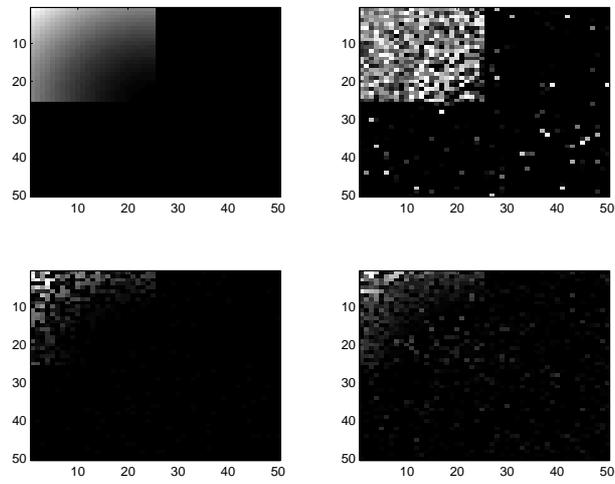}
\caption{Top, left: non-linearity coefficients $\gamma_{1,2}$. Top,
right: distance between $\Ubar$ and $\hat{\U}_n$ estimated with
$\eta=0$. Bottom: distance between $\bar{\U}$ and $\hat{\U}_\ell$
estimated with $\eta=0.5$ (left) and $\eta=50$ (right).}
\label{fig:dist_map_L=2_knn=3_spect_nb_synth_GBM_nb}
\end{figure}

First, we investigate the estimation of the subspace $\mathcal{U}$
when the image pixels are non-linear functions of the abundances. For this purpose, a $50
\times 50$ synthetic hyperspectral image is generated following a
recently introduced non-linear model referred to as generalized bilinear model
(GBM). As indicated in \cite{Halimi2011}, the GBM is notably well
adapted to describe non-linearities due to multipath effects. It
assumes that the observed pixel spectrum $\y_\ell$ can be written
\begin{equation} \label{eq:GBM}
\y_\ell = \sum_{r=1}^{R} a_{r,\ell} \bm_r + \sum_{i=1}^{R-1}\sum_{j=i+1}^{R} \gamma_{i,j,\ell} a_{i,\ell}a_{j,\ell} \bm_i \odot \bm_j
\end{equation}
where $\odot$ stands for the Hadamard (termwise) product and the
abundances $a_{r,\ell}$ ($r=1,\ldots,R$) satisfy the constraints in
\eqref{eq:constraints}. In \eqref{eq:GBM}, the parameters
$\gamma_{i,j,\ell}$ (which belong to $[0,1]$) characterize the
importance of non-linear interactions between the endmembers $\bm_i$
and $\bm_j$ in the $\ell$-th pixel. In particular, when
$\gamma_{i,j,\ell}=0$ ($\forall i,j$), the GBM reduces to the
standard LMM \eqref{eq:LMM}. Moreover, when $\gamma_{i,j,\ell}=1$
($\forall i,j$), the GBM leads to the non-linear model introduced by
Fan \emph{et al.} in \cite{Fan2009}. In this simulation, the
synthetic image has been generated using the GBM with $R=3$
endmember signatures extracted from a spectral library. The
corresponding abundances have been uniformly drawn in the set
defined by the constraints \eqref{eq:constraints}. We have assumed
that there is no interaction between endmembers $\bm_1$ and $\bm_3$,
and between endmembers $\bm_2$ and $\bm_3$ resulting in
$\gamma_{1,3,\ell}=\gamma_{2,3,\ell}=0$, $\forall \ell$.
Moreover, the interactions between endmembers $\bm_1$ and $\bm_2$
are defined by the map of coefficients $\gamma_{1,2,\ell}$ displayed
in Fig. \ref{fig:dist_map_L=2_knn=3_spect_nb_synth_GBM_nb}~(top,
left panel) where a black (resp. white) pixel represents the lowest
(resp. highest) degree of non-linearity.  As can be seen in this
figure, $75\%$ of the pixels (located in the bottom and upper right
squares of the image) are mixed according to the LMM resulting in
$\gamma_{1,2,\ell}=0$. The $25\%$ remaining image pixels (located in
the upper left square of the image) are mixed according to the GBM
with nonlinearity coefficients $\gamma_{1,2,\ell}$ radially
increasing from $0$ to $1$ ($\gamma_{1,2,\ell}=0$ in the image
center and $\gamma_{1,2,\ell}=1$ in the upper left corner of the
image). Note that this image contains a majority of pixels that are
mixed linearly and belong to a common subspace of $\mathbb{R}^2$.
Conversely, the non-linearly mixed pixels do not belong to this
subspace\footnote{Assuming there is a majority of image pixels that
are mixed linearly is a reasonable assumption for most hyperspectral
images.}. We propose here to estimate the local subspace
$\mathcal{U}_\ell$ where a given image pixel $\y_\ell$ and its
nearest spectral neighbors ${\mathcal{V}_\ell^{(K-1)}}$ live
(${\mathcal{V}_\ell^{(K-1)}}$ denotes the set of the ($K-1$)-nearest
neighbors of $\y_\ell$).

Assuming as a first approximation that all the image pixels are linearly mixed, all these pixels are approximately contained in a common $2$-dimensional subspace $\bar{\mathcal{U}}$ that can be determined by performing a PCA of $\y_1,\ldots,\y_L$ (see \cite{Dobigeon2009sp} for more details). The corresponding principal vectors spanning $\bar{\mathcal{U}}$ are gathered in a matrix $\Ubar$. This matrix $\Ubar$ is used as \emph{a priori} knowledge regarding the $2$-dimensional subspace containing $\left\{\y_\ell,{\mathcal{V}_\ell^{(K-1)}}\right\}_{\ell=1,\ldots,L}$. However, this crude estimation can be refined by the Bayesian estimation strategy developed in the previous sections. More
precisely, for each pixel $\y_\ell$, we compute the MMSD estimator
of the $N\times p$ matrix $\U_\ell$, whose columns are supposed to
span the subspace $\mathcal{U}_\ell$ containing $\y_\ell$ and its
$K-1$-nearest neighbors ${\mathcal{V}_\ell^{(K-1)}}$. The Bayesian
estimator $\hat{\U}_\ell$ is computed from its closed-form
expression \eqref{Ummsd_LM_Bingham}, i.e., using the Bingham prior
where $\Ubar$ has been introduced above. Then, for each pixel, we
evaluate the distance between the two projection matrices
$\hat{\U}_{\ell} \hat{\U}_{\ell}^{T}$ and $\Ubar \Ubar^{T}$ onto the subspaces
$\hat{\mathcal{U}}_\ell = \range{\hat{\U}_\ell}$ and
$\bar{\mathcal{U}} = \range{\Ubar}$, respectively. As stated in
Section~\ref{section:MMSD}, the natural distance between these two
projection matrices is given by
$d^2\left(\hat{\U}_\ell,\Ubar\right)=2\left(p-\Tr{\hat{\U}_\ell^T\Ubar\Ubar^T\hat{\U}_\ell}\right)$.
The resulting distance maps are depicted in Fig.
\ref{fig:dist_map_L=2_knn=3_spect_nb_synth_GBM_nb} (bottom panels)
for $2$ non-zero values of $\eta
\triangleq 2\varn \kappa$ (as it can be noticed in
\eqref{Ummsd_LM_Bingham}, this hyperparameter $\eta$ balances the
quantity of \emph{a priori} knowledge $\Ubar$ included in the
estimation with respect to the information brought by the data).
For comparison purpose, the subspace $\hat{\mathcal{U}}_\ell$ has
been also estimated by a crude SVD of
$\left\{\y_\ell,\mathcal{V}_\ell^{(K-1)}\right\}$ (top right panel).
In this case, $\hat{\U}_\ell$ simply reduces to the associated
principal singular vectors and can be considered as the MMSD
estimator of $\U_\ell$ obtained for $\eta=0$.

These figures show that, for the $75\%$ of the pixels generated
using the LMM (bottom and right parts of the image), the subspace
$\bar{\mathcal{U}}$ estimated by an SVD of the whole dataset
$\y_1,\ldots,\y_L$ is very close to the hyperplanes
$\hat{\mathcal{U}}_\ell$ locally estimated from
$\left\{\y_\ell,\mathcal{V}_\ell^{(K-1)}\right\}$ through the
proposed approach (for any value of $\eta$). Regarding the remaining $25\%$ pixels resulting from the
GBM (top left part of the image), the following comments can be
made. When a crude SVD of $\left\{\y_\ell,\mathcal{V}_\ell^{(K-1)}\right\}$ is conducted,
i.e., when no prior knowledge is taken into account to compute the MMSD ($\eta=0$, top right panel), the distance between the locally estimated subspace
$\hat{\mathcal{U}}_\ell$ and the \emph{a priori} assumed hyperplane
$\bar{\mathcal{U}}$ does not reflect the non-linearities contained
in the image. Conversely, when this crude SVD is regularized by
incorporating prior knowledge with
$\eta=0.5$ and $\eta=50$ (bottom left and right panels,
respectively), leading to the MMSD estimator, the larger the degree
of non-linearity, the larger the distance between $\Ubar$ and
$\hat{\U}_\ell$. To summarize, evaluating the distance between the
MMSD estimator $\hat{\U}_\ell$ and the \emph{a priori} given matrix
$\Ubar$ allows the degree of non-linearity to be quantified. This
interesting property is exploited on a real hyperspectral image in
the following section.

\subsection{Real data}
\begin{figure}[p]
\centering
\includegraphics[width=8cm]{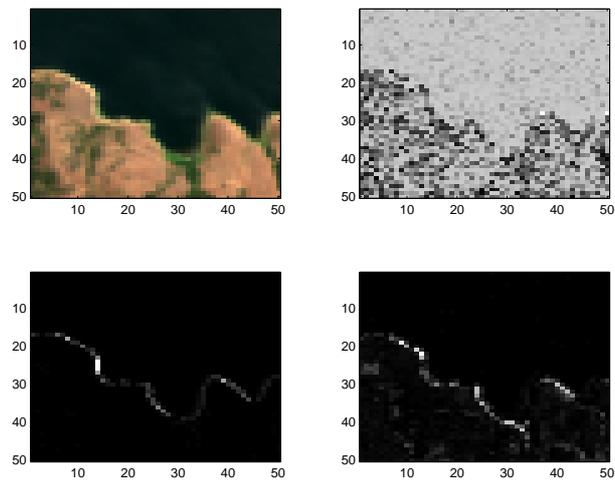}
\caption{Top, left: The Moffett Field scene as composite true
colors. Top, right: distance between $\Ubar$ and $\hat{\U}_n$
estimated with $\eta=0$. Bottom: distance between $\bar{\U}$ and
$\hat{\U}_\ell$ estimated with $\eta=0.5$ (left) and $\eta=50$
(right).} \label{fig:dist_map_L=2_knn=3_spect_nb}
\end{figure}

The real hyperspectral image considered in this section has been
acquired in 1997 over Moffett Field, CA, by the NASA spectro-imager
AVIRIS. This image, depicted with composite true colors in Fig.
\ref{fig:dist_map_L=2_knn=3_spect_nb} (top, left panel), has been
minutely studied in \cite{Dobigeon2009sp} assuming a linear mixing
model. The scene consists of a large part of a lake (black pixels,
top) and a coastal area (bottom) composed of soil (brown pixels) and
vegetation (green pixels), leading to $R=3$ endmembers whose spectra
and abundance maps can be found in \cite{Dobigeon2009sp}. A simple
estimation of a lower-dimensional space $\bar{\mathcal{U}}$ where
the pixels live can be conducted through a direct SVD of the whole
dataset, providing the \emph{a priori} matrix $\Ubar$. As
in the previous section, this crude estimation can be refined by
computing locally the MMSD estimators $\hat{\U}_{\ell}$ spanning the
subspaces $\hat{\mathcal{U}}_\ell$ (bottom panels). These estimators
have been also computed with $\eta=0$, corresponding to an SVD of
$\left\{\y_\ell,\mathcal{V}_\ell^{(K-1)}\right\}$ (top, right
figure). The distances between $\Ubar$ and $\hat{\U}_\ell$ have been
reported in the maps of Fig. \ref{fig:dist_map_L=2_knn=3_spect_nb}.
Again, for $\eta=0$ (top, right panel), a simple local SVD is unable to locate possible
non-linearities in the scene. However, for two\footnote{Additional results obtained with other values of $\eta$ are available online at
\url{http://dobigeon.perso.enseeiht.fr/app_MMSD.html}.} non-zero values $\eta=0.5$ and $\eta=50$
(bottom left and right panels, respectively), the distances between
the \emph{a priori} recovered subspace $\bar{\mathcal{U}}$ and the
MMSD-based subspace $\hat{\mathcal{U}}_\ell$ clearly indicate that
some non-linear effects occur in specific parts of the image,
especially in the lake shore. Note that the non-linearities
identified by the proposed algorithm are very similar to the ones
highlighted in \cite{Halimi2011} where the unmixing procedure was
conducted by using the GBM defined in \eqref{eq:GBM}. This shows the
accuracy of the proposed MMSD estimator to localize the
non-linearities occurring in the scene, which is interesting for the
analysis of hyperspectral images.

\section{Conclusions}
This paper considered the problem of estimating a subspace using
some available a priori information. Towards this end, a Bayesian
framework was advocated, where the subspace $\U$ is assumed to be
drawn from an appropriate prior distribution. However, since we
operate in a Grassmann manifold, the conventional MMSE approach is
questionable as it amounts to minimizing a distance which is not the
most meaningful on the Grassmann manifold. Consequently, we
revisited the MMSE approach and proposed, as an alternative, to
minimize a natural distance on the Grassmann manifold. A general
framework was formulated resulting in a novel estimator which
entails computing the principal eigenvectors of the posterior mean
of $\U \U^{T}$. The theory was exemplified on a few simple examples,
where the MMSD estimator can either be obtained in closed-form or requires resorting to an MCMC simulation method. The new
approach enables one to combine efficiently the prior knowledge
and the data information, resulting in a method that performs well
at low SNR or with very small sample support. A successful application to the analysis of non-linearities contained in hyperspectral images was also presented.

\appendices
\section{The eigenvalue decomposition of $\int \U \U^T p_{\mathrm{B}} (\U) d \U$ \label{app:eigBingham}}
The purpose of this appendix is to prove the following proposition which can be invoked to obtain the MMSD estimator whenever the posterior distribution $p(\U | \Y)$ is a Bingham distribution.

\begin{prop} \label{prop:eigBingham} Let $\U \in \mathbb{R}^{N \times p}$ be an orthogonal matrix -$\U^{T}\U=\I$- drawn from a Bingham distribution with parameter matrix $\A$
\begin{equation}\label{p_B(A)}
p_{\mathrm{B}} (\U)  =  \ex{- \kappa_{\mathrm{B}}(\A)} \etr{\U^{T} \A \U }
\end{equation}
with $\kappa_{\mathrm{B}}(\A)= \ln \oneFone{\frac{1}{2}p}{\frac{1}{2}N}{\A}$. Let $\A=\U_a \bLambda_a \U_a^T$ denote the eigenvalue decomposition of $\A$ where the eigenvalues are ordered in descending order. Let us define $\M = \int \U \U^{T} p_{\mathrm{B}} (\U) d\U$. Then the eigenvalue decomposition of $\M$ writes
\begin{equation*}
\M = \ex{- \kappa_{\mathrm{B}}(\A)} \U_a \bGamma \U_a^T
\end{equation*}
with $\bGamma=\frac{\partial \ex{\kappa_{\mathrm{B}}(\A)}}{\partial \bLambda_a}$ and $\gamma_1 \geq \gamma_2 \geq \cdots \geq \gamma_N$ where $\gamma_n=\bGamma(n,n)$.
\end{prop}
\begin{IEEEproof}
For notational convenience, let us work with the projection matrix $\bP=\U\U^T$ whose distribution on the Grassmann manifold is \cite{Chikuse03}
\begin{equation}\label{p(P)}
p(\bP)  =  \ex{- \kappa_{\mathrm{B}}(\A)} \etr{\bP \A}.
\end{equation}
We have then that
\begin{align*}
\M &= \ex{- \kappa_{\mathrm{B}}(\A)} \int \bP \etr{\bP \U_a \bLambda_a \U_a^T} d \bP \\
&= \ex{- \kappa_{\mathrm{B}}(\A)} \U_a \left[ \int \U_a^T \bP \U_a \etr{\U_a^T \bP \U_a \bLambda_a} d \bP \right] \U_a^T  \\
&= \ex{- \kappa_{\mathrm{B}}(\A)} \U_a \left[ \int \bP \etr{\bP \bLambda_a} d \bP \right] \U_a^T  \\
&= \ex{- \kappa_{\mathrm{B}}(\A)} \U_a \bGamma \U_a^T.
\end{align*}
Moreover $\bGamma$ is diagonal since, for any orthogonal diagonal matrix $\D$,
\begin{align*}
\bGamma \D &=  \int \bP \D \etr{\bP \bLambda_a} d \bP \\
&= \D \left[ \int \D^T \bP \D \etr{\D^T \bP \D \D^T \bLambda_a \D} d \bP \right] \\
&=  \D \int \bP \etr{\bP \bLambda_a} d \bP \\ &= \D \bGamma
\end{align*}
where, to obtain the third line, we made use of the fact that $\D^T \bLambda_a \D=\bLambda_a$. It follows that the eigenvectors of $\M$ and $\A$ coincide, and that the eigenvalues of $\M$ are $\ex{- \kappa_{\mathrm{B}}(\A)} \gamma_n$, for $n=1,\cdots,N$.
Moreover, it is known that $\ex{- \kappa_{\mathrm{B}}(\A)}=\ex{- \kappa_{\mathrm{B}}(\bLambda_a)}$ and, from \eqref{p(P)}, one has
\begin{equation*}
\ex{\kappa_{\mathrm{B}}(\bLambda_a)} =  \int \etr{\bP \bLambda_a} d \bP.
\end{equation*}
Differentiating the latter equation with respect to $\lambda_a(k)$ and denoting $p_n=\bP(n,n)$, one obtains
\begin{align*}
\frac{\partial \ex{\kappa_{\mathrm{B}}(\bLambda_a)} }{\partial \lambda_a(k)} &=
\frac{\partial}{\partial \lambda_a(k)} \int \ex{\sum_{n=1}^{N} \lambda_a(n) p_n} d \bP \\
&=\int p_k \etr{\bP \bLambda_a} d \bP \\ &= \gamma_k.
\end{align*}
The previous equation enables one to relate the eigenvalues of $\A$ and those of $\M$. It remains to prove that $\gamma_1 \geq \gamma_2 \geq \cdots \geq \gamma_N$. Towards this end, we make use of a very general theorem due to Letac \cite{Letac10}, which is briefly outlined below. Let
\begin{equation*}
P(\mu,\A)(d\X) = \ex{\kappa_{\mu}(\A)}\etr{\X^T \A} \mu(d\X)
\end{equation*}
be a probability associated with a unitarily invariant measure $\mu$ on the set of $N \times N$ symmetric matrices. Consider the case of a diagonal matrix $\A=\diag{a_1,a_2,\cdots,a_N}$ with $a_1 \geq a_2 \geq \cdots \geq a_N$. Then \cite{Letac10} proves that $\M=\int \X P(\mu,\A)(d\X)$ is also diagonal, and moreover if $\M=\diag{m_1,m_2,\cdots,m_N}$ then $m_1 \geq m_2 \geq \cdots \geq m_N$. Use of this theorem completes the proof of the proposition.
\end{IEEEproof}
\begin{rem}
Most of the proposition could be proved, however in a rather indirect way, using the results in \cite{Jupp79}. In this reference, Jupp and Mardia consider maximum likelihood estimation of the parameter matrix $\A$ from the observation of $K$ independent matrices $\U_k$ drawn from \eqref{p_B(A)}. Let $\Pbar=K^{-1} \sum_{k=1}^{K} \U_k \U_k^T$ and let its eigenvalue decomposition be $\Pbar = \Vbar \Dbar \Vbar^T$. Then the maximum likelihood estimate $\hat{\A}$ of $\A$ has eigenvalue decomposition $\hat{\A}=\Vbar \D\Vbar^T$ with $\bar{d}_n= \partial \ex{\kappa_{\mathrm{B}}(\D)} / \partial d_n$. Moreover, due to Barndorff-Nielsen theorem for exponential families, one has
\begin{equation*}
\tau_{B}(\hat{\A}) = \int \bP \ex{\kappa_{B}(\hat{\A})} \etr{\bP \hat{\A}}  d\bP=\Vbar \Dbar \Vbar^T
\end{equation*}
which proves, since $\hat{\A}=\Vbar \D\Vbar^T$, that
\begin{equation*}
\int \bP \ex{\kappa_{B}(\D)} \etr{\bP \Vbar \D\Vbar^T}  d\bP = \Vbar \Dbar \Vbar^T.
\end{equation*}
In \cite{Jupp79} however, no results about the ordering of the eigenvalues was given.
\end{rem}

\section{Sampling from the Bingham-von Mises Fisher distribution \label{app:BMF}}
In this appendix, we show how to sample a unitary random matrix $\X
\in \mathbb{R}^{N \times p}$ from a (matrix) Bingham von Mises
Fisher (BMF) distribution, $\X \sim \BMF{\A}{\B}{\C}$. As will be
explained shortly, this amounts to sampling successively each column
of $\X$, and entails generating a random unit norm vector drawn from
a (vector) BMF distribution. We briefly review how to sample the
columns of $\X$ and then explain how to sample from a vector BMF
distribution.
\subsection{The matrix BMF distribution}
The density of $\X \sim \BMF{\A}{\B}{\C}$ is given by
\begin{align}\label{matrix_BMF}
p\left( \X | \A, \B, \C \right) & \propto \etr{\C^{T} \X + \B \X^{T} \A \X} \nonumber \\
&\propto \prod_{k=1}^{p} \ex{\bc_{k}^{T} \x_k + \B(k,k) \x_k^T  \A \x_k}
\end{align}
where $\X=\begin{bmatrix}\x_1 & \x_2 & \cdots & \x_p \end{bmatrix}$ and  $\C=\begin{bmatrix}\bc_1 & \bc_2 & \cdots & \bc_p \end{bmatrix}$. In \cite{Hoff09} a Gibbs-sampling strategy was presented in order to sample from this distribution, in the case where $\A$ is full-rank. We consider here a situation where $\A$ is rank-deficient and therefore we need to bring appropriate modifications to the scheme of \cite{Hoff09} in order to handle the rank deficiency of $\A$. As evidenced from \eqref{matrix_BMF} the distribution of $\A$ is a product of vector BMF distributions, except that the columns of $\X$ are not statistically independent since they are orthogonal with probability one. Let us rewrite $\X$ as $\X = \begin{bmatrix} \x_1 & \cdots & \x_{k-1} & \Null \z & \x_{k+1} & \cdots & \x_{p} \end{bmatrix}$ where $\z \in \mathcal{S}_{N-p+1}= \left\{ \x \in \mathbb{R}^{N-p+1 \times 1} ; \x^T \x =1 \right\}$ and $\Null$ is an $N \times N-p+1$ orthonormal basis for $\range{\X_{-k}}^{\perp}$ where $\X_{-k}$ stands for the matrix $\X$ with its $k$-th column removed. As shown in \cite{Hoff09} the conditional density of $\z$ given $\X_{-k}$ is
\begin{align}
p \left( \z | \X_{-k} \right) &\propto \ex{\bc_{k}^{T} \Null \z + \B(k,k) \z^{T} \Null^{T} \A \Null \z } \nonumber \\
&\propto \ex{\tilde{\bc}_{k}^{T} \z + \z^{T} \tilde{\A} \z }
\end{align}
where $\tilde{\bc}_{k}=\Null^{T} \bc_{k}$ and $\tilde{\A}=\B(k,k) \Null^{T} \A \Null$. Therefore, $\z | \X_{-k}$ follows a vector BMF distribution $\z | \X_{-k} \sim \vBMF{\tilde{\A}}{\tilde{\bc}_{k}}$. A Markov chain that converges to $\BMF{\A}{\B}{\C}$ can thus be constructed as follows:
\begin{algorithmic}[1]
\REQUIRE initial value $\X^{(0)}$
\FOR{$k=1,\cdots,p$ (random order)}
\STATE compute a basis $\Null$ for the null space of $\X_{-k}$ and set $\z= \Null^{T} \x_k$.
\STATE compute $\tilde{\bc}_{k}=\Null^{T} \bc_{k}$ and $\tilde{\A}= \B(k,k) \Null^{T} \A \Null$.
\STATE sample $\z$ from a $ \vBMF{\tilde{\A}}{\tilde{\bc}_{k}}$ distribution (\emph{see next section}). \label{sample_vBMF}
\STATE set $\x_k= \Null \z$.
\ENDFOR
\end{algorithmic}
\subsection{The vector BMF distribution}
The core part of the above algorithm, see line \ref{sample_vBMF}, is
to draw a unit-norm random vector $\x$ distributed according to a
vector Bingham-von Mises Fisher distribution. The latter
distribution on the $M$-dimensional sphere has a density with
respect to the uniform distribution given by
\begin{equation}
p\left( \x | \bc, \A \right) \propto \ex{\bc^{T} \x + \x^{T} \A \x}, \x \in \mathcal{S}_{M}.
\end{equation}
In \cite{Hoff09} a Gibbs-sampling strategy
was presented in order to sample from this distribution. While $\A$
was assumed to be full-rank in \cite{Hoff09}, we consider here a
situation where $\A$ is rank-deficient, i.e. its eigenvalue
decomposition can be written as $\A = \bE \bLambda \bE^{T}$ where
$\bE$ stands for the orthonormal matrix of the eigenvectors and
$\bLambda = \diag{\lambda_1,\lambda_2,\cdots,\lambda_r,0,\cdots,0}$
is the diagonal matrix of its eigenvalues. Our derivation follows
along the same lines as in \cite{Hoff09} with the appropriate
modifications due to the rank deficiency of $\A$. Let $\y=\bE^ {T}\x
\in \mathcal{S}_{M}$ and $\bd = \bE^{T} \bc$. Since $y_M^2 = 1-
\sum_{k=1}^{M-1} y_k^2$, the uniform density in terms of the
unconstrained coordinates $\left\{ y_1,y_2,\cdots,y_{M-1} \right\}$
is proportional to $|y_M|^{-1}$ and the density of $\left\{
y_1,y_2,\cdots,y_{M-1} \right\}$ is given by \cite{Hoff09}
\begin{align}
p\left( \y | \bd, \bE \right) &\propto \ex{\bd^{T} \y + \y^{T} \bLambda \y} |y_M|^{-1} , \quad y_M^2 = 1- \sum_{k=1}^{M-1} y_k^2 \nonumber \\  &\propto \ex{\sum_{k=1}^{M} d_k y_k + \sum_{k=1}^{r} \lambda_k y_k^2} |y_M|^{-1}.
\end{align}
In order to sample from this distribution, a Gibbs sampling strategy
is advocated. Towards this end, we need to derive the conditional
distributions of $y_k$, given $\y_{-k}$ where $\y_{-k}$ stands for
the vector $\y$ with its $k$-th component removed. Similarly to
\cite{Hoff09}, let us make the change of variables $\theta_k=y_k^2$
and let $\q = \begin{bmatrix} \frac{y_1^2}{1-y_k^2} &
\frac{y_2^2}{1-y_k^2} & \cdots & \frac{y_{M}^2}{1-y_k^2}
\end{bmatrix}^{T}$, so that $\left\{y_1^2,y_2^2,\cdots,y_M^2\right\}
= \left\{ \theta_k, \left(1-\theta_k\right) \q_{-k} \right\}$. Since
this change of variables is not bijective, i.e. $y_k \pm
\theta_k^{1/2}$, we need to introduce the sign $s_k$ of $y_k$, and
we let $\bs = \begin{bmatrix} s_1 & s_2 & \cdots & s_M
\end{bmatrix}^{T}$. Note that $y_M^2 = 1-\sum_{k=1}^{M-1} y_k^2$,
$|y_M|=(1-\theta_{k})^{1/2}q_M^{1/2}$ and $q_M=1-\sum_{\ell=1,\ell
\neq k}^{M-1} q_{\ell}$. As shown in \cite{Hoff09}, the Jacobian of
the transformation from  $\left\{y_1,y_2,\cdots,y_{M-1}\right\}$ to
$\left\{ \theta,q_1,\cdots,q_{k-1},q_{k+1},\cdots,q_{M-1} \right\}$
is proportional to $\theta_k^{-1/2}
\left(1-\theta_k\right)^{(M-2)/2} \prod_{\ell=1,\ell \neq k}^{M-1}
q_{\ell}^{-1/2}$, and therefore the joint distribution of $ \theta_k
, s_k , \q_{-k} , \bs_{-k}$ can be written as
\begin{align}\label{joint_pdf_theta_q}
p\left( \theta_k , s_k , \q_{-k} , \bs_{-k} \right) &\propto \theta_k^{-1/2} \left(1-\theta_k\right)^{(M-3)/2} \left(  \prod_{\ell \neq k} q_{\ell}^{-1/2} \right) \nonumber \\
&\times \ex{s_k \theta_k^{1/2} d_k + \left(1-\theta_k\right)^{1/2} \sum_{\ell \neq k} d_{\ell} s_{\ell} q_{\ell}^{1/2}} \nonumber \\
& \times \begin{cases} \ex{\theta_k \lambda_k + \left(1-\theta_k\right) \sum_{\ell=1,\ell \neq k}^{r} q_{\ell} \lambda_{\ell}} & 1 \leq k \leq r \\  \ex{\left(1-\theta_k\right) \sum_{\ell=1}^{r} q_{\ell} \lambda_{\ell}} & r+1 \leq k \leq M \end{cases}.
\end{align}
It follows that
\begin{itemize}
\item for $k\in \left[1,r\right]$
\begin{align}
p\left( \theta_k , s_k | \q_{-k} , \bs_{-k} \right) &\propto \theta_k^{-1/2} \left(1-\theta_k\right)^{(M-3)/2} \ex{\theta_k \lambda_k + \left(1-\theta_k\right) \q_{-k}^{T} \blambda_{-k}} \nonumber \\
& \times  \ex{s_k \theta_k^{1/2} d_k + \left(1-\theta_k\right)^{1/2} \left[ \bs_{-k} \odot \q_{-k}^{1/2} \right]^{T} \bd_{-k}}.
\end{align}
\item for $k\in \left[r+1,M\right]$
\begin{align}
p\left( \theta_k , s_k | \q_{-k} , \bs_{-k} \right) &\propto \theta_k^{-1/2} \left(1-\theta_k\right)^{(M-3)/2} \ex{\left(1-\theta_k\right) \q^{T} \blambda} \nonumber \\
& \times  \ex{s_k \theta_k^{1/2} d_k + \left(1-\theta_k\right)^{1/2} \left[ \bs_{-k} \odot \q_{-k}^{1/2} \right]^{T} \bd_{-k}}.
\end{align}
\end{itemize}
In the previous equations, $\odot$ stands for the element-wise
vector or matrix product and $\q_{-k}^{1/2}$ is a short-hand
notation to designate the vector $\begin{bmatrix} q_{1}^{1/2} &
\cdots & q_{k-1}^{1/2} & q_{k+1}^{1/2} & \cdots & q_{M}^{1/2}
\end{bmatrix}^{T}$. In order to sample from $p\left( \theta_k , s_k
| \q_{-k} , \bs_{-k} \right)$, we first sample $\theta_k$ from
\begin{align}\label{pdf_thetak_cond}
p\left( \theta_k | \q_{-k} , \bs_{-k} \right) &= p\left( \theta_k , s_k=-1 | \q_{-k} , \bs_{-k} \right) + p\left( \theta_k , s_k=1 | \q_{-k} , \bs_{-k} \right) \nonumber \\
&\propto \theta_k^{-1/2} \left(1-\theta_k\right)^{(M-3)/2} \ex{a_k \theta_k + b_k \left(1-\theta_k\right)^{1/2}} \nonumber \\
&\times \left[ \ex{-d_k \theta_k^{1/2}} + \ex{-d_k \theta_k^{1/2}} \right]
\end{align}
where $b_k = \left[ \bs_{-k} \odot \q_{-k}^{1/2} \right]^{T} \bd_{-k}$ and
\begin{equation}
a_k = \begin{cases} \lambda_k - \q_{-k}^{T} \blambda_{-k} & k \in \left[1,r\right] \\
-\q^{T} \blambda & k\in \left[r+1,M\right] \end{cases}.
\end{equation}
Next, we sample $s_k \in \left\{-1,+1\right\}$ with probabilities
proportional to $\left(e^{-d_k \theta_k^{1/2}}, e^{+d_k
\theta_k^{1/2}} \right)$. In order to sample from the distribution
in \eqref{pdf_thetak_cond}, an efficient rejection sampling scheme
was proposed in \cite{Hoff09}, where the proposal distribution is a
beta distribution with suitably chosen parameters.

\section*{Acknowledgment} The authors would like to thank Prof. Kit Bigham from the University of Minnesota for insightful comments on the Bingham distribution and for pointing reference \cite{Jupp79}. They are also indebted to Prof. G\'erard Letac, University of Toulouse, for fruitful discussions leading to the proof of Proposition \ref{prop:eigBingham} given in Appendix \ref{app:eigBingham}.

\bibliographystyle{IEEEtran}
\bibliography{mmsd}

\newpage




\end{document}